\documentclass[12pt]{article}
\usepackage[utf8]{inputenc}
\usepackage{authblk}
\usepackage{setspace}
\usepackage[margin=2.0cm]{geometry}
\usepackage{graphicx}
\graphicspath{ {./figures/} }
\usepackage{amsmath}
\usepackage{amsfonts}
\usepackage{lineno}
\usepackage{hyperref}
\usepackage[document]{ragged2e}
\usepackage{subcaption}

\usepackage{algorithm}
\usepackage{algpseudocode}

\usepackage{etoolbox}
\AtBeginEnvironment{figure}{\setlength{\RaggedRightParindent}{0em}}
\AtBeginEnvironment{table}{\setlength{\RaggedRightParindent}{0em}}

\usepackage[style=authortitle,
            citestyle=authoryear,
            backend=biber,
            natbib=true,
            maxnames=2,
            minnames=1,
            maxbibnames=3,
            minbibnames=3,
            giveninits=true,
            terseinits=true,
            sorting=nyt]{biblatex}

\AtEveryCitekey{\renewcommand*{\mkbibnamegiven}[1]{}}

\DeclareNameFormat{family-given}{%
  \namepartfamily
  \ifblank{\namepartgiveni}{}{\addspace\namepartgiveni}%
  \ifthenelse{\value{listcount}<\value{liststop}}
    {\addcomma\addspace}
    {}%
  \usebibmacro{name:andothers}}

\DefineBibliographyStrings{english}{
  andothers = {
    \ifbibliography
      {\mkbibemph{et\addabbrvspace al\adddot}}
      {et\addabbrvspace al\adddot}
  }
}

\DeclareNameAlias{author}{family-given}
\DeclareNameAlias{editor}{family-given}
\DeclareNameAlias{translator}{family-given}

\AtBeginBibliography{%
}

\DeclareFieldFormat{year}{#1}
\DeclareFieldFormat{date}{#1}

\renewbibmacro{in:}{}

\DeclareFieldFormat[article]{title}{#1}

\renewbibmacro*{journal+issuetitle}{
  \usebibmacro{journal}
  \setunit*{\addspace}
  \usebibmacro{issue+date}
  \setunit{\addcolon\space}
  \usebibmacro{issue}
  \newunit}

\renewbibmacro*{issue+date}{
  \printfield{year}
  \setunit{\addcomma\space}
}

\title{A method for empirically assessing small area estimators via bootstrap-weighted k-Nearest-Neighbor artificial populations, with applications to forest inventory}
\date{} 
\author[1,2,*]{Grayson W. White} 
\author[3]{Jerzy A. Wieczorek} 
\author[3]{Zachariah W. Cody}
\author[3]{Emily X. Tan}
\author[3]{Jacqueline O. Chistolini}
\author[4]{Kelly S. McConville}
\author[5]{Tracey S. Frescino}
\author[5]{Gretchen G. Moisen}
\affil[1]{{\small Department of Mathematics \& Statistics, Reed College, Portland, OR, USA}}
\affil[2]{{\small Department of Forestry, Michigan State University, East Lansing, MI, USA}}
\affil[3]{{\small Department of Statistics, Colby College, Waterville, ME, USA}}
\affil[4]{{\small Dominguez Center for Data Science, Bucknell University, Lewisburg, PA, USA}}
\affil[5]{{\small Rocky Mountain Research Station, USDA Forest Service, Riverdale, UT, USA}}
\affil[*]{{\small Corresponding author: Grayson W. White. Contact:\newline gwhite@reed.edu, 3203 Woodstock Blvd, Portland, OR 97202}}

\begin{document}
\maketitle

\newpage

\begin{abstract}
National Forest Inventories (NFIs) monitor forest attributes across a variety of spatial and temporal scales in a given country. Increased interest in reporting and management at smaller scales has driven NFIs to investigate and adopt small area estimation (SAE) due to the promise of increased precision at these scales. However, comparing and evaluating SAE models for a given application is inherently difficult. Typically, many areas lack enough data to check unit-level modeling assumptions or to assess unit-level predictions empirically; and no ground truth is available for checking area-level estimates. Design-based simulation from artificial populations can help with each of these issues, but only if the artificial populations realistically represent the application at hand and are not built using assumptions that inherently favor one SAE model over another. In this paper, we borrow ideas from random hot deck, approximate Bayesian bootstrap (ABB), and $k$ Nearest Neighbor (kNN) imputation methods to propose a kNN-based approximation to ABB (KBAABB), for generating an artificial population when rich unit-level auxiliary data is available. We introduce diagnostic checks on the process of building the artificial population, and we demonstrate how to use such an artificial population for design-based simulation studies to compare and evaluate SAE models, using real data from the United States Department of Agriculture, Forest Service, Forest Inventory and Analysis Program, the NFI of the United States. 
\end{abstract}


\newpage

\section{Introduction}

In recent years, National Forest Inventories (NFIs) have been tasked with providing precise estimates of forest attributes at progressively finer spatial and temporal scales. For example, in the United States, the USDA 2014 and 2018 Farm Bills \citep{usda2014farm,usda2018farm} direct the Forest Inventory and Analysis (FIA) program (\url{https://www.fs.usda.gov/research/programs/fia}) to: ``implement procedures to improve the statistical precision of estimates at the sub-State level'' and ``find efficiencies in the operations of the Forest Inventory and Analysis Program under section 3(e) of the Forest and Rangeland Renewable Resources Research Act of 1978 (16 U.S.C. 1642(e)) through the improved use and integration of advanced remote sensing technologies to provide estimates for State- and national-level inventories, where appropriate.'' The FIA program provides data and tools for delivering standardized estimates at state levels over 5-10 year periods, but there is currently no standardized methodology to generate estimates at these sub-State levels or shorter time intervals. In response to the Farm Bill initiatives, FIA is working towards a strategic plan to define FIA user needs, evaluate small area estimation (SAE) methodology, and deliver statistically sound, standardized estimates to users at the sub-State level.

Motivation for producing estimates at sub-State levels extends beyond congressional directives, as producing estimates of forest attributes at these finer scales is of the utmost importance for management and ecological monitoring given global climate change and our dependence on natural resources. Specifically, the need for more precise information regarding forest attributes at sub-State levels is essential for field specialists and managers to make tactical assessments for local and broad-scale monitoring and planning. With these needs in mind, combined with the availability of high quality auxiliary data from satellite imagery, climate records, elevation models, etc., scientists and managers alike are turning to SAE methods, which are designed to address the limitations of small sample sizes in unplanned domains, to better understand forest landscapes across the globe. 

The increased interest in and use of SAE methods from NFIs and their users has the potential for substantial benefits to forest management and monitoring. However, much caution must be taken when utilizing SAE methods, especially those of the model-based variety. This is because model-based methods do not have the same statistical properties as many design-based methods (e.g. design-unbiasedness). Furthermore, caution must be taken when implementing SAE methods from the design-based paradigm that only exhibit asymptotic statistical properties (e.g. asymptotic design-unbiasedness) such as model-assisted methods including the post-stratification and generalized regression estimators. For model-based estimators, and for design-based estimators with only asymptotic statistical guarantees, it is imperative to empirically evaluate estimators to ensure accurate and unbiased estimation. SAE methods may produce estimates and estimated uncertainty intervals that appear precise for some forest parameter of interest in a small region, but these estimates are only reliable if the underlying model has met the necessary modeling assumptions. Thus, without sufficient evaluation of modeling assumptions, SAE methods can lead decision-makers astray. Further, countless SAE methods exist for estimating a particular forest parameter of interest in a particular small area, but understanding the best SAE method to use for each particular NFI application is still an open question. Furthermore, empirically comparing and evaluating SAE models is an inherently challenging problem \citep{brown2001evaluation,rao15small,dorfman2018towards,tzavidis2018start}. While there exist many useful model-comparison tools for unit-level predictive models (for instance, cross-validation, or information criteria such as AIC and BIC), these tools are often of limited use for SAE. Sometimes we do not have unit-level data, only area-level. Even when we do have unit-level data, what we really want to assess are the area-level estimates, not the unit-level predictions themselves. In both cases, area-level direct estimates from the sample itself are not precise enough to be treated as ground truth for model comparisons; we cannot rely on checking whether the model-based estimate is close to the direct estimate for each area.

\citet{dorfman2018towards} lists a ``Variety of Inadequate Methods of Evaluation'' for SAE, including ``(7) large scale \emph{simulation studies} from administrative, census or large samples---these can give useful insights but satisfactory extrapolation to the case at hand has to be assumed[\ldots]''
From the design-based perspective, such simulation studies are most straightforward to justify if they consist of repeated sampling from a real and relevant population \citep{lehtonen2009design, tzavidis2018start}, but a complete real population is rarely available. Alternately, simulation studies can consist of repeated sampling from an \emph{artificial} population \citep{alfons2011simulation, wieczorek2012bayesian, wieczorek2013empirical, templ2017simulation}, but this requires extra care to justify why the artificial population resembles the real one in the relevant sense: How can we trust that model assessment across samples drawn from the artificial population should be informative about model performance on our \emph{real} sample from the \emph{real} population?

In the present paper, we describe a novel
approach to generating one or more artificial populations: a ``kNN-based approximation to the approximate Bayesian bootstrap'' (KBAABB). We justify why KBAABB can allow for ``satisfactory extrapolation'' from simulation studies. We illustrate how to use our artificial population for model evaluation and comparison, following approaches and metrics similar to those in \citet{dorfman2018towards} and \citet{wieczorek2012bayesian}.
The objectives of this study were: (1) to develop a novel nonparametric method for generating artificial populations, KBAABB; (2) to demonstrate empirical diagnostics for assessing the properties of artificial populations generated by KBAABB; and (3) to illustrate how KBAABB can be used to compare SAE models on a real dataset.

Section~\ref{sec:Data} introduces the specific FIA data and auxiliary variables used for our analyses. Briefly, in our setting we have complete auxiliary data, denoted $X$, for the entire population of interest, the M333 Northern Rocky Mountain Forest-Steppe-Coniferous Forest-Alpine Meadow Province, but a much smaller survey sample of the joint auxiliary and response variables at each FIA plot, which we denote $(X,Y)$. 
To generate a complete artificial population, we wish to impute response values from the observed survey data (donor dataset) to every row of the full-population auxiliary data (recipient dataset). 
We introduce methodology to generate the artificial population in Section~\ref{sec:KBAABB}. We begin with the approximate Bayesian bootstrap (ABB), a multiple imputation tool that approximates the process of drawing from a posterior distribution for the missing data given the observed data \citep{rubin1986multiple}. We use bootstrap-weighted sampling with kNN to implement KBAABB, a computationally-cheaper approximation of ABB. The resulting artificial population can be interpreted as one draw from a posterior distribution for the population's response values. Alternatives to KBAABB are discussed below in this Introduction

A single run of KBAABB produces one artificial population, from which we repeatedly draw samples using the FIA's quasi-systematic sampling design. This allows us to carry out design-based simulations in a finite-population framework. In addition, repeated runs of KBAABB could easily produce multiple artificial populations, allowing for simulations under a superpopulation framework \citep{isaki1982survey} and/or Bayesian posterior inference, although we do not do so in this paper. In Section~\ref{sec:Diagnostics} we assess the qualities of the artificial population generated with KBAABB from FIA data. 
Empirically, we show that the output of KBAABB on our FIA dataset is 
a reasonable population to compare small area estimators on. Further, in Supplementary material~\ref*{sec:Sensitivity} we show the artificial population is comparable to or better than artificial populations built via other imputation methods commonly used in practice: single nearest neighbor (NN) imputation, and unweighted kNN imputation where each donor is selected uniformly at random from a donor pool of the recipient's $k$ closest neighbors \citep{andridge2023adapting}. KBAABB avoids the worst-case risks of single NN and the \emph{ad hoc} nature of unweighted kNN. 

Our artificial population is designed to have the correct marginal distribution of $X$ and a reasonable estimate of the conditional distributions of each $Y|X$, and our samples from the artificial population accurately represent the real sampling design. However, we do not claim that the area-level means from the artificial population are correct for the real population. Rather, we will use the artificial population to seek SAE methods that tend to produce good estimates---close to the ``artificial truth'' of the artificial population's area-level means---when fitted to samples drawn from the artificial population. To demonstrate the use of our methodology in practice, Section~\ref{sec:CaseStudies} evaluates a set of common SAE models for estimation of average basal area per acre on the design-based samples taken from the artificial population generated with KBAABB for M333. We assess estimator quality from insights taken from the KBAABB artificial population, finding substantial differences in estimator performance in this FIA setting. This provides a valuable example of the use case of KBAABB in practice for forest managers, SAE practitioners, ecologists, and other like-minded researchers. Finally, we conclude our study with Section~\ref{sec:discussion}, a discussion of the implications of our findings, and future work.

By contrast with KBAABB,
other work on realistic artificial populations for design-based simulations typically relies only on the survey dataset itself, or on the survey data combined with population summaries of the auxiliary variables rather than a complete unit-level dataset.
\citet{alfons2011simulation} generate fully-synthetic artificial populations through a combination of weighted sampling from the real survey data and model-based simulation from parametric models fit to the real survey data. \citet{wieczorek2013empirical} treat a large sample as the full artificial population and mimic the real sampling design to take smaller subsamples from it for the purpose of evaluating SAE models. 
Beyond the survey dataset alone, if full-population marginal totals or cross-tabulations for the auxiliary data are also available, e.g.\ tables from a recent census, \citet{templ2017simulation} summarize approaches to generating an artificial population by resampling entire rows from the survey data until the artificial population meets these marginal or cross-tabulated constraints.
However, none of the above approaches make use of the rich unit-level auxiliary data that is available in our FIA setting.

Another approach is to make up a parametric model (or fit one to the sample); make a prediction $\hat Y(X_i)$ for each row of the $X$s; and add random noise to each prediction to generate the artificial population's $Y$s \citep{morris2014tuning}.
However, under this approach, the artificial population would inherently favor SAE models of the same parametric form. If some of the models we wish to evaluate are correct for the artificial population but mis-specified for the real population, then simulation studies on the artificial population would be overly optimistic for these models.
Instead, KBAABB's nonparametric kNN-based matching ensures that the artificial population will not have a bias towards being fit well by any of the common parametric SAE models, yet our imputed $Y$ values will still have realistic conditional distributions for $Y|X$. Of course, if we also planned to evaluate the use of kNN-based models for small area estimators, then we would want to choose a different non-kNN model to generate our artificial population  \citep{baffetta2009design}.

\section{Methods}\label{sec:Methods}

\subsection{Data}\label{sec:Data}

\begin{figure}
    \includegraphics[width=.9\textwidth]{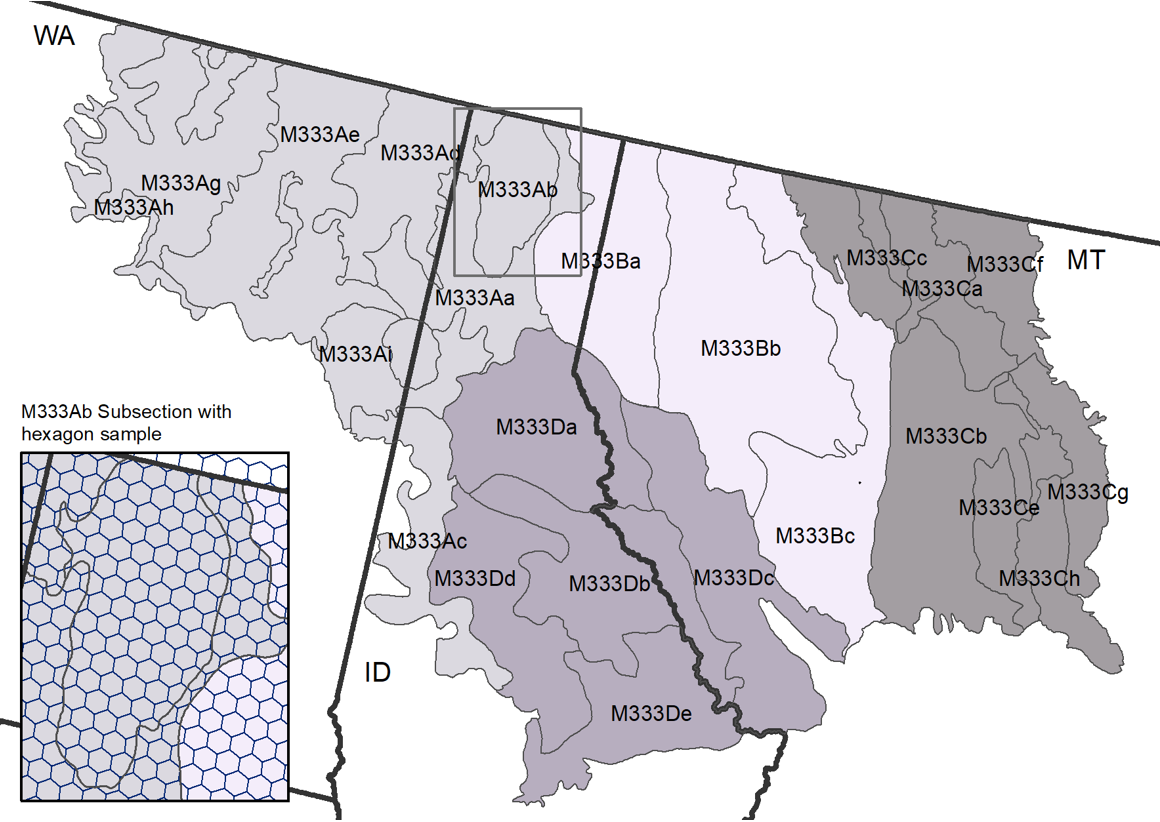}
    \caption{Simulation study area.}
    \label{fig:M333HexSample}
\end{figure}

Our simulation study area is the M333 Northern Rocky Mountain Forest-Steppe-Coniferous Forest-Alpine Meadow Province from the US Forest Service's national hierarchical framework of ecological units \citep{cleland2007ecological, ecomap1993national}. EcoMap provinces are delineated by climatic zones and broad vegetation types with further delineations into sections and subsections, representing different geomorphic and topographic features \citep{ecomap1993national}. The M333 province stretches across western Montana, through Northern Idaho, to eastern Washington and is about 98,700 km$^2$ in size. The province encompasses four mountainous sections, with an average of six subsections each, twenty-three in total. These are our 23 domains (Figure~\ref{fig:M333HexSample}).

The FIA program surveys a sample of field plot locations across the US based on a hexagonal grid. Each plot samples approximately 2.5 acres (1.0 ha) of land and represents approximately 6000 acres (2428 ha) of land within each hexagon \citep{mcroberts2005enhanced, bechtold2005enhanced}. Attributes are measured and observed at each survey plot location and stored in FIA's national database \citep{burrill2021forest}. We extracted data from FIA's database (last updated 2021 July 8) on 20 July 2021. There are 3946 survey plots across the 23 domains, with 98\% of the plots measured between 2010 and 2019. From these plots, we compile our Y-variables and associated X-variables.

Our Y-variables were all aggregated to the plot level and calculated on a per-acre basis.
These variables were: basal area (\texttt{BA}), net cubic-foot volume (\texttt{VOLCFNET}), net board-foot volume (\texttt{VOLBFNET}), number of trees (\texttt{COUNT}), and aboveground dry biomass (\texttt{DRYBIO}). The \texttt{VOLCFNET} response variable measures the average net volume of trees per acre, where the \texttt{VOLBFNET} response variable measures the average net volume of usable timber per acre. See Table~\ref{tab:YVariables} for further details.

\begin{table}
 \centering
 \caption{Y-variables or survey variables recorded by field crews at each surveyed plot.}
 \label{tab:YVariables}
 \footnotesize
 \begin{tabular}{ l|l|l } 
 Variable & Units & Description \\ 
 \hline
 \texttt{BA} & square feet / acre & Basal area of live trees \\ 
 \texttt{VOLCFNET} & cubic feet / acre & Net cubic-foot volume in sawlog portion of a live sawtimber tree \\
 \texttt{VOLBFNET} & board feet / acre & Net board-foot volume in sawlog portion of a live sawtimber tree \\
 \texttt{COUNT} & number of trees / acre & Number of live trees per acre unadjusted \\
 \texttt{DRYBIO} & tons / acre & Aboveground biomass of live trees \citep{heath2009investigation}
 \end{tabular}
\end{table}

We also have auxiliary X-variables from satellites, climate records, and digital elevation models that were thought to relate to and discriminate our Y variables (Table~\ref{tab:XVariables}). The auxiliary information includes surface reflectance images, including spectral indices and classified imagery to get a picture of vegetation on the ground \citep{schwager2021remote} and both broad-scale and localized climatic characteristics to understand ecological tolerances.
All auxiliary data were resampled from their original resolution to 90x90m pixels. For each plot, we have the auxiliary data corresponding to that plot’s location.
Although more auxiliary variables could have been used, our kNN matching steps only used the categorical variable \texttt{wc2cl} for stratification and the quantitative variables in  Table~\ref{tab:XVariables}
for matching. Some of our analyses in Supplementary material~\ref*{sec:Shiny} used \texttt{tnt} for post-stratification.

\begin{table}
 \centering
 \caption{X-variables or auxiliary variables.}
 \label{tab:XVariables}
 \footnotesize
 \begin{tabular}{ l|l|l } 
 Variable & Units & Description \\ 
 \hline
 \texttt{tcc} & \% & National Land Cover Dataset (NLCD) Analytical Tree Canopy Cover \citep{yang2018new} \\ 
 \texttt{elev} & meters & LANDFIRE 2010 DEM - elevation \citep{usgs2019ned} \\  
 \texttt{eastness} & (-100 to 100) & Transformed aspect degrees to eastness \citep{usgs2019ned} \\ 
 \texttt{northness} & (-100 to 100) & Transformed aspect degrees to northness \citep{usgs2019ned} \\ 
 \texttt{tri} & index & Terrain Ruggedness Index \citep{usgs2019ned} \\ 
 \texttt{tpi} & index & Topographic Position Index \citep{usgs2019ned} \\ 
 \texttt{ppt} & mm$\times$100 & PRISM mean annual precipitation - 30yr normals (1991-2020) \citep{daly2002knowledge} \\ 
 \texttt{tmin01} & $^{\circ}$C$\times$100 & PRISM mean minimum temperature (Jan) - 30yr normals \citep{daly2002knowledge} \\ 
 \texttt{wc2cl} & class & Eur.\ Space Agency (ESA) 2020 WorldCover global land cover \citep{zanaga2021esa} \\ 
 \texttt{tnt} & class & LANDFIRE 2014 tree/non-tree lifeform mask \citep{rollins2009landfire, picotte2019landfire}
 \end{tabular}
\end{table}

As a basis for the artificial population, besides the survey sample we also have complete auxiliary-only data for the full population: 11,752,067 pixels in the 4286 hexagons that fall within or overlap with EcoProvince M333. The largest hexagon had 2979 pixels, and the median was 2964, although several hexagons along the edge of the province had just a few pixels in the target population. 
For each of these auxiliary-only pixels, we had the same $X$ variables as for the 3946 pixels corresponding to the real survey plots.

We use an analysis function in the \texttt{FIESTAnalysis} R package (which is an extension of the \texttt{FIESTA} R package), \texttt{anGetPixeldat()}, to generate pixel-level data tables for each subsection \citep{frescino2022fiestanalysis, frescino2023fiesta}. The function creates a table of a combination of longitude and latitude extracted from each 90x90m pixel across each subsection, including a unique identifier of each pixel. This table is then used to extract values of our auxiliary information to create a set of data for our simulation models. 

\subsection{Implementation of KBAABB}\label{sec:KBAABB}

\subsubsection{Artificial population generation}

With rich auxiliary data for the entire population, 
we can treat the unobserved response values in our population as missing data that needs to be imputed.
Several popular approaches to imputation of missing data are based on nearest neighbors (NN). In the single NN approach, each recipient row is matched to the ``nearest'' donor row---the donor row whose auxiliary variables are most similar to the recipient's by some distance metric---and that donor's response variables are imputed to that recipient. However, 
this approach risks overusing the same donor for all recipients with similar $X$ values, which could result in unrealistically low variability in the imputed $Y$ values \citep{andridge2023adapting}. For the same reason, imputing the means or medians of larger donor pools \citep{crookston2008yaimpute} is also unsuitable for our purposes.

An alternative is to use a recipient's $k$ nearest neighbors (kNN) as the donor pool, for a $k$ such as 5, 10, or 20 \citep{morris2014tuning}, and then to choose one case in this donor pool \emph{uniformly at random} to be imputed to the recipient \citep{andridge2021finding}. This avoids single NN's problem of always imputing the same donor---but it can also lead to bias in the imputed conditional distributions of $Y|X$, since a larger donor pool may include $Y$ values that are not realistic for the recipient's $X$ values. It also requires careful choice of $k$. (In the extreme case, $k=n$ would replace every imputed conditional distribution of $Y|X$ with an imputed marginal distribution of $Y$.)

Our proposed method, KBAABB, works like kNN imputation but with uniform random sampling of donors replaced by weighted sampling, with the weights based on bootstrap selection probabilities. The $j^{th}$ NN is selected with probability $p_b \times (1-p_b)^{j-1}$, where $p_b = 1-e^{-1} \approx 0.632$ is the asymptotic probability of an observation being selected into a bootstrap sample \citep{andridge2023adapting}.
Since these probabilities become negligible quickly, in practice we only find each recipient's $k=10$ NNs to use as its donor pool, then use bootstrap-weighted sampling from this pool to select one donor for that recipient.
See inner loop of Algorithm~\ref{alg:alg1}.

Furthermore, to reduce bias and variance, KBAABB can be applied independently within strata; and to avoid sensitivity to units of measurement, the auxiliary variables can be centered and scaled before calculating distances to find the $k$ NNs for each recipient. See outer loop of Algorithm~\ref{alg:alg1}. By default we use Euclidean distance; alternatives are discussed in Section~\ref{sec:Choices}. 

If there are several response $Y$ variables to be imputed for each recipient, KBAABB imputes all of them from the same donor to their recipient. If instead we were to impute each response variable independently, our artificial population would not accurately reflect the associations among these variables \citep{andridge2010review,nur2005dealing}.

\begin{algorithm}
\caption{KBAABB, with stratification and standardization}
\begin{algorithmic}
    \State{$k \gets 10$} 
    \Comment{\texttt{\emph{Choose $k$ and distance metric, or use defaults}}}
    \State{\texttt{DistMetric} $\gets$ \text{Euclidean distance}}
    \\
    \\
    \For{$j=1$ to $k-1$} 
    \Comment{\texttt{\emph{Define the selection probabilities}}}
        \State{$\mathbb{P}\left(j\text{th ranked NN selected}\right) \gets \exp{(-j+1)} \times (1 - \exp{(-1)}) $}
    \EndFor
    \State{$\mathbb{P}\left(k\text{th ranked NN selected}\right) \gets 1 - \sum_{j=1}^{k-1}{\mathbb{P}\left(j\text{th ranked NN selected}\right)} $}
    \\
    \For{\texttt{stratum} $\in\{\texttt{strata}\}$}  \Comment{\texttt{\emph{Process each stratum separately}}}
        \State{\texttt{recipients} $\gets \{\text{all population units} \in \texttt{stratum}\} $}
        \State{\texttt{donors} $\gets \{\text{all sample units} \in \texttt{stratum}\}$}
        \State{Center/scale auxiliary data in \texttt{recipients} and \texttt{donors}, using \texttt{recipients} means and SDs}
        \\
            \For{$\texttt{unit} \in \texttt{recipients}$} \Comment{\texttt{\emph{kNN matching and ABB imputation}}}
                \State{$\texttt{Dists} \gets$ distances to \texttt{unit} from all \texttt{donors}, via \texttt{DistMetric} in auxiliary data space}
                \State{$\texttt{NNs} \gets$ $k$ nearest neighbors to \texttt{unit} from \texttt{donors}, according to \texttt{Dists}}
                \State{Rank \texttt{NNs} by distance to \texttt{unit} in auxiliary data space from $j = 1$ to $k$}
                \State{Randomly sample one element of \texttt{NNs} based on $\mathbb{P}\left(j\text{th ranked NN selected}\right)$}
                \State{Impute all response values from selected element of \texttt{NNs}  to \texttt{unit}}
            \EndFor
    \EndFor
\end{algorithmic}
\label{alg:alg1}
\end{algorithm}

In principle, by selecting nearer neighbors with higher probability, KBAABB should avoid the pitfalls of single NN (too little variability in $Y$) and kNN (bias due to large $k$), while also removing the need to choose tuning parameter $k$.
Empirically, in Supplementary material~\ref*{sec:Sensitivity}, we report a sensitivity analysis regarding the probabilities of selection and the choice of $k$, comparing KBAABB to unweighted kNN approaches on our FIA dataset. For each response variable $Y$, we found the same results: As kNN's $k$ grows, the distribution of Y in each domain becomes more similar across domains, and unweighted $k = 20$ or larger makes the domains too similar to each other. Furthermore, when looking at spatial maps of the imputed $Y$ values, the spatial patterns look realistically smooth for single NN and for KBAABB, but begin to look unrealistically noisy for unweighted $k = 5$ or larger.

\subsubsection{Heuristic rationale for bootstrap weighting}

KBAABB's use of bootstrap weights is heuristically inspired by the approximate Bayesian bootstrap (ABB) \citep{rubin1986multiple}. ABB is popular for multiple imputation because it can be proven to be ``valid,'' a technical term in the imputation literature, loosely meaning that it allows asymptotically unbiased design-based inferences under both the sampling and the nonresponse mechanisms; for details, see Chapter 4 of \citet{rubin1987multiple}. In Bayesian inference, one run of ABB can also be treated as one draw from a posterior distribution for the population's response values, given the survey sample and the full-population auxiliary values and using a nonparametric model for the likelihood. Neither ``validity'' nor Bayesian inference are central to our goal of creating an artificial population, but ABB inspired us to try bootstrap weighting, and empirically we have found it to be a success in our application; see Supplementary material~\ref*{sec:Sensitivity}.

Briefly, our bootstrap weights arise from mimicking true ABB, where we would first draw a bootstrap sample from the complete-data cases, then draw with replacement from this bootstrap sample to impute the missing cases. In order to reduce both bias and variance, typically ABB imputes from a smaller donor pool based on some auxiliary information.
If we use a single NN from the bootstrap sample to define the donor pool for each recipient, 
around $p_b$ of recipients will have their NN from the original sample selected as their donor \citep{andridge2023adapting}.
However, NN with true ABB is still computationally expensive for our task of generating an artificial population, as we would need to take a new bootstrap sample of the complete-case survey data for every recipient, then redo the NN search for that recipient. KBAABB is a computational shortcut: Use the original sample (not a resample) to find a donor pool of $k$ NNs for each recipient row separately, then use bootstrap-weighted sampling to choose one donor for that recipient. Under true ABB with NN, each recipient has approximately a $(1-p_b)^{j-1}\times p_b$ probability that the first $j-1$ NNs would not be in the bootstrap sample but the $j^{th}$ would be. (This probability becomes negligible quickly, so in practice we only find the $k=10$ NNs and select the $10^{th}$ NN with probability $1 - \sum_{j=1}^{9}(1-p_b)^{j-1}\times p_b \approx 0.000124$.) Thus, KBAABB's bootstrap weights approximate the process of running NN with true ABB once per recipient.

\subsubsection{Considerations in implementation}\label{sec:Choices}

To implement KBAABB, as with any other kNN-based method we must decide how to define ``nearest'' for the kNN search: Which variables, transformations, and distance metric should be used to rank the donor pool for each recipient? For a detailed discussion of how we chose to define ``nearest'' for our study, see Supplementary material~\ref*{sec:Implementation}. We now discuss the considerations to be made in general.

First, if the population consists of strata or post-strata for which units tend to be homogeneous within strata but quite different across strata, KBAABB will have less bias and less unnecessary variability if kNN matching is carried out separately in each stratum.

Next, kNN is known to suffer from the curse of dimensionality: if we use Euclidean distance in a high-dimensional space, the several nearest neighbors might all be very far away \citep{james2021introduction}. When there are many auxiliary variables available for matching,
NN imputation can be asymptotically biased and some analysts prefer to summarize the auxiliary variables into a univariate matching variable \citep{morris2014tuning, yang2019nearest}.
However, to keep the matching process as model-free as possible, others may prefer to simply reduce the matching space to a subset of the variables, using subject matter expertise. Such parsimony can also reduce unnecessary variance in the matching process \citep{dorazio2015integration}.

Furthermore, if matching variables are heavily skewed, it may be worthwhile to transform them to be more symmetric. Otherwise, donors with distant outlier values might never be one of the nearest neighbors for any recipient. \citet{chen2000nearest} show that for imputation purposes, NN imputation can be biased when the matching variables are not symmetric. 

We may also center and scale each of the matching variables to have mean 0 and standard deviation 1 (within each stratum separately) on the full-population auxiliary dataset. The same centering and scaling constants (by class and variable) should be applied to the smaller survey sample dataset. This removes the effect of different variables having different units and scales, ensuring that each variable is given equivalent weight in the kNN matching process \citep{james2021introduction}.

Finally, a distance metric must be chosen and computed between each recipient and each donor using the selected and transformed variables. We use Euclidean distance as the KBAABB default and as the distance for the case study in Section~\ref{sec:CaseStudies}. Euclidean distance should suffice when the selected variables are not strongly correlated with each other and they have been centered and scaled. However, if necessary in other applications, alternatives such as e.g.\ Mahalanobis distance could account for covariances between variables \citep{crookston2008yaimpute}.

\subsubsection{Sampling for design-based simulations}

For each replication (``rep'') of a survey sample from the KBAABB-generated artificial population, we take a sample using the same sampling design that was used to draw the real sample. 
This lets us carry out design-based simulations which respect the sampling design of the actual survey data. When we fit a given SAE model form to each rep, and then summarize the model's empirical properties across reps, these properties are reasonable stand-ins for what we could expect to see under repeated sampling from the real sampling design.
For our application, we took design-based samples using FIA's quasi-systematic sampling design (see Supplementary material~\ref*{sec:Sampling}).

\subsection{Case study: estimators and performance metrics}\label{sec:CaseStudiesMethods}

We have chosen four commonly applied estimators in the SAE literature to assess in M333, to give a sense of what can be done with KBAABB through a basic yet illustrative example, with results discussed in Section~\ref{sec:CaseStudies}. In particular, we implemented two design-based estimators, the Horvitz-Thompson \citep{horvitz52HT} and modified generalized regression (GREG) \citep{woodruff1966use, wojcik2022gregory}, along with two model-based estimators, the area-level empirical best linear unbiased prediction (EBLUP) Fay-Herriot \citep{fay1979estimates} and the unit-level EBLUP Battese-Harter-Fuller \citep{battese1988error}. Model-based and -assisted estimators used auxiliary variables \texttt{tcc}, \texttt{tri}, and \texttt{elev}, which are defined in Table~\ref{tab:XVariables}. 

These estimators were implemented in R using the \texttt{FIESTAutils} function \texttt{SAest.large()} \citep{frescino2023fiestautils}. The Horvitz-Thompson estimator was fit without weights and hence its estimate and variance estimate are the sample mean and sample variance, respectively. The modified GREG estimator's regression coefficients were fit with ordinary least squares and the estimate and variance estimate were created based on the standard equations described in \citet{rao15small}. The Fay-Herriot and Battese-Harter-Fuller estimators were fit with restricted maximum likelihood to ensure unbiased estimation of the random effects. For further details regarding the estimators and their statistical properties, see \citet{rao15small}. 

We considered four evaluation metrics in this case study: relative bias, MSE, MSE ratio, and 95\% confidence interval coverage. The relative bias is defined as, for a particular estimator ($i$) and domain ($j$),
\[
    \frac{\mathbb{E}[\hat \mu_{ij}] - \mu_{j}}{\mu_{j}}
\]
where $\mu_{j}$ is the true average basal area per acre in subsection $j$ as generated by KBAABB and $\mathbb{E}[\hat \mu_{ij}]$ is defined as follows,
\begin{equation} \label{eq:empirical_expectation}
    \mathbb{E}[\hat \mu_{ij}] = \frac{1}{K} \sum_{k=1}^K \hat\mu_{ijk},
\end{equation}
i.e., the empirical expected value for estimator $i$ of average basal area per acre in subsection $j$ across all reps ($k$ from 1 to $K$). The MSE for a given estimator and subsection is given by
\[
    \mbox{MSE}_{ij} = \frac{1}{K} \sum_{k=1}^K \left( \hat\mu_{ijk} - \mu_j \right)^2.
\]
The MSE ratio is defined as the ratio of the empirical expected value of the MSE estimator and the empirical MSE for a particular estimator and area of interest, 
\[
    \frac{\mathbb{E}[\widehat{\mbox{MSE}}_{ij}]}{\mbox{MSE}_{ij}}
\]
where $\mathbb{E}[\widehat{\mbox{MSE}}_{ij}]$ is the empirical expected value of the MSE estimator for estimator $i$ in subsection $j$ across all $K$ reps, defined analogously to Equation~\eqref{eq:empirical_expectation}. The 95\% confidence interval coverage is defined as, for a particular estimator and subsection, the proportion of samples across all reps where a given estimator's 95\% confidence interval contains $\mu_{ij}$.

\section{Results}

We now turn to discussing the results of generating our artificial population and the comparison of small area estimators in estimation of average basal area per acre in Sections~\ref{sec:Diagnostics} and \ref{sec:CaseStudies}, respectively. We discuss the results and diagnostics of the artificial population generation in order to properly evaluate that our artificial population is well-suited for evaluating estimators. 

\subsection{Empirical diagnostics for KBAABB in our FIA setting}\label{sec:Diagnostics}

To check whether the artificial population generated with KBAABB looked reasonable, we compared the statistical distributions of the imputed Y variables to the original survey.
We also looked for imbalances in how often each donor was used, and in how often each recipient's domain matched the donor's domain. We repeated these checks for each of the Y-variables and found similar results, so for brevity most examples in this section are illustrated using a single Y variable: average basal area per acre.

\begin{figure}
    \includegraphics[width=.49\textwidth]{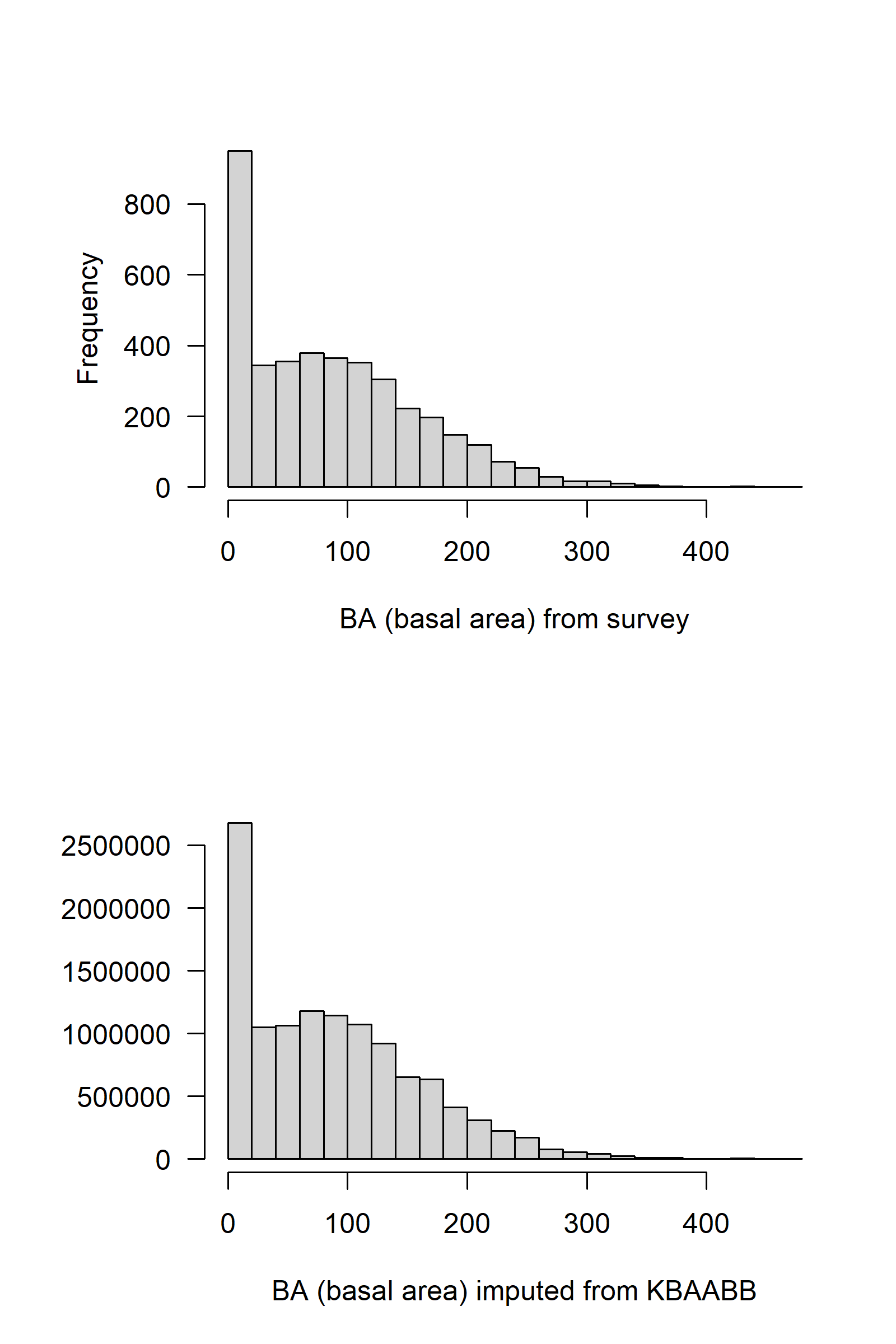}
    \includegraphics[width=.49\textwidth]{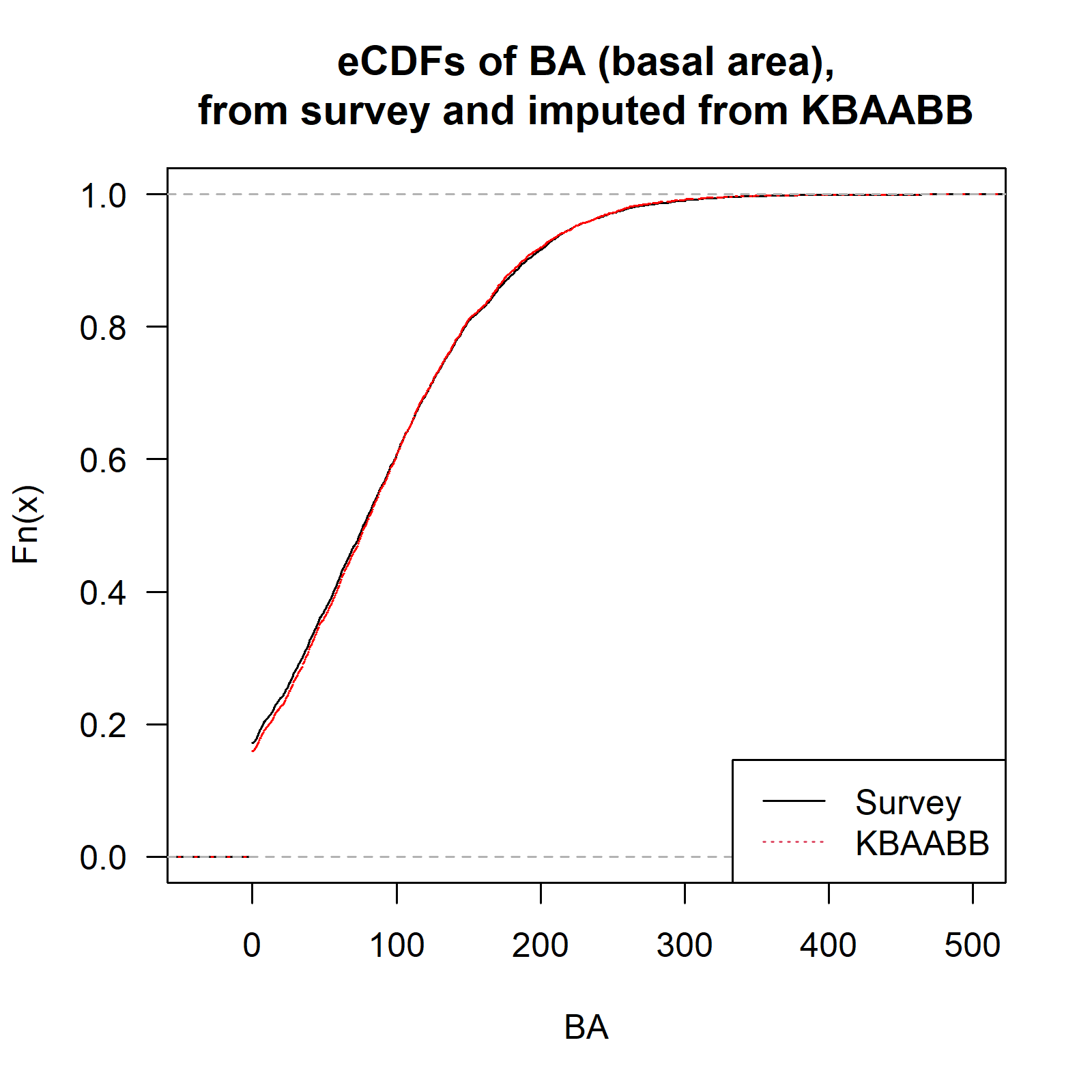}
    \caption{Histograms (left) and empirical CDFs (right) of original vs imputed average basal area per acre.}
    \label{fig:Hists_BA}
\end{figure}

First, we confirmed that marginal distributions 
from the artificial population resembled the real survey data by comparing the original and imputed values of each variable using histograms or eCDFs (empirical cumulative distribution functions), as in Figure~\ref{fig:Hists_BA}.
Histograms are more familiar to many readers, but eCDFs do not require us to choose a bin size and can also make it easier to compare behavior in the tails. With either graph type, we found that each Y variable's marginal distribution did not change substantially between the survey and the artificial population. 

\begin{figure}
    \includegraphics[width=\textwidth]{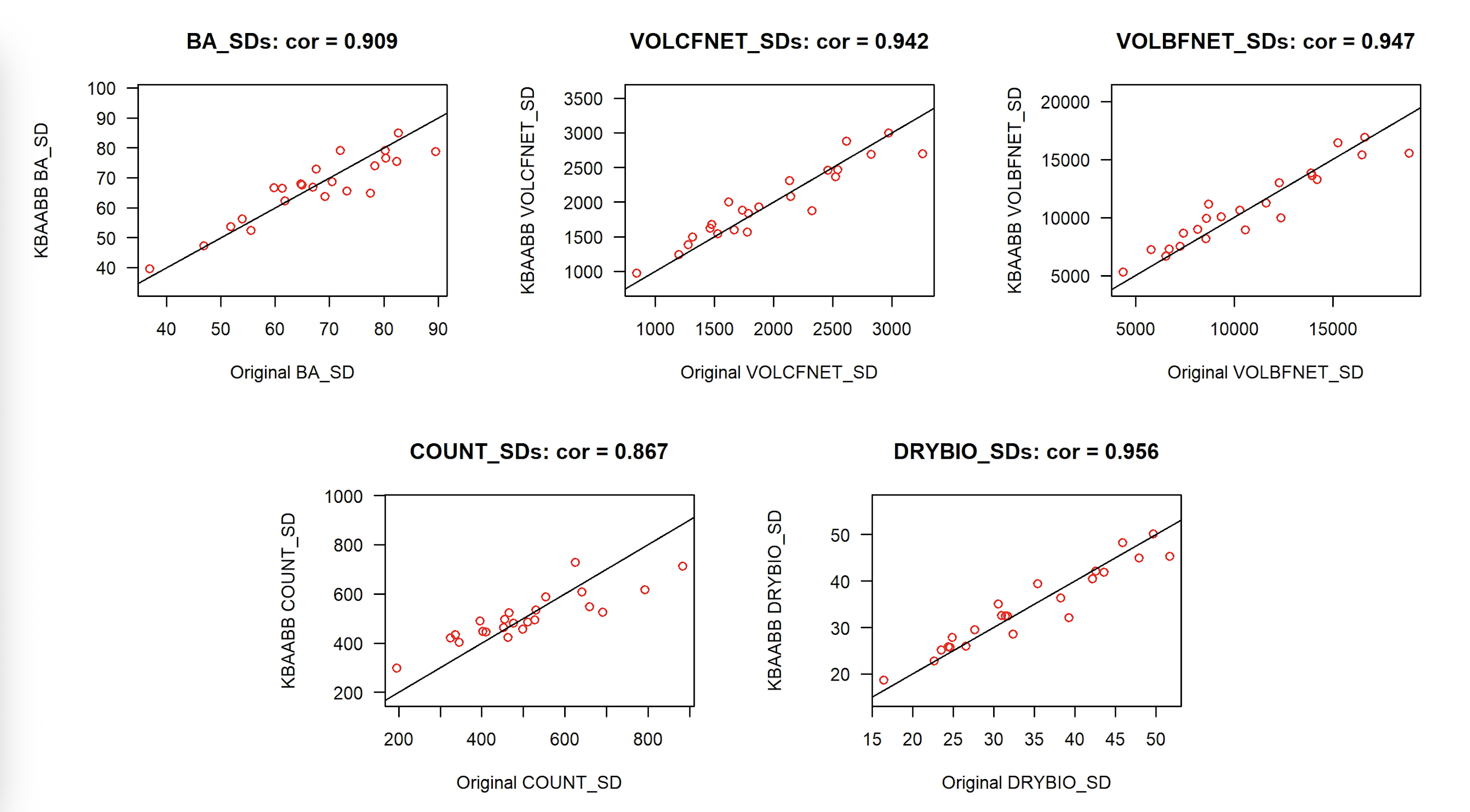}
    \caption{Original vs imputed SDs of each Y-variable for each subsection, using KBAABB. All plots show $y=x$ line as a common reference.}
    \label{fig:SDs_KBAABB}
\end{figure}

Next, we checked how well the imputed domain-level standard deviations (SDs) match the real survey SDs in each domain. Figure~\ref{fig:SDs_KBAABB} shows that for each Y-variable, the KBAABB artificial population's domain-level SDs are strongly but not perfectly correlated with their original-sample counterparts. This is a good sign, because if they were perfectly correlated we might simply be reproducing the original sample, while if they had low correlations we might be imputing unrealistic donors to each recipient. For a similar example using kNN with varying values of $k$, please see Supplementary material~\ref*{sec:Sensitivity}.

\begin{figure}
    \includegraphics[width=.45\textwidth]{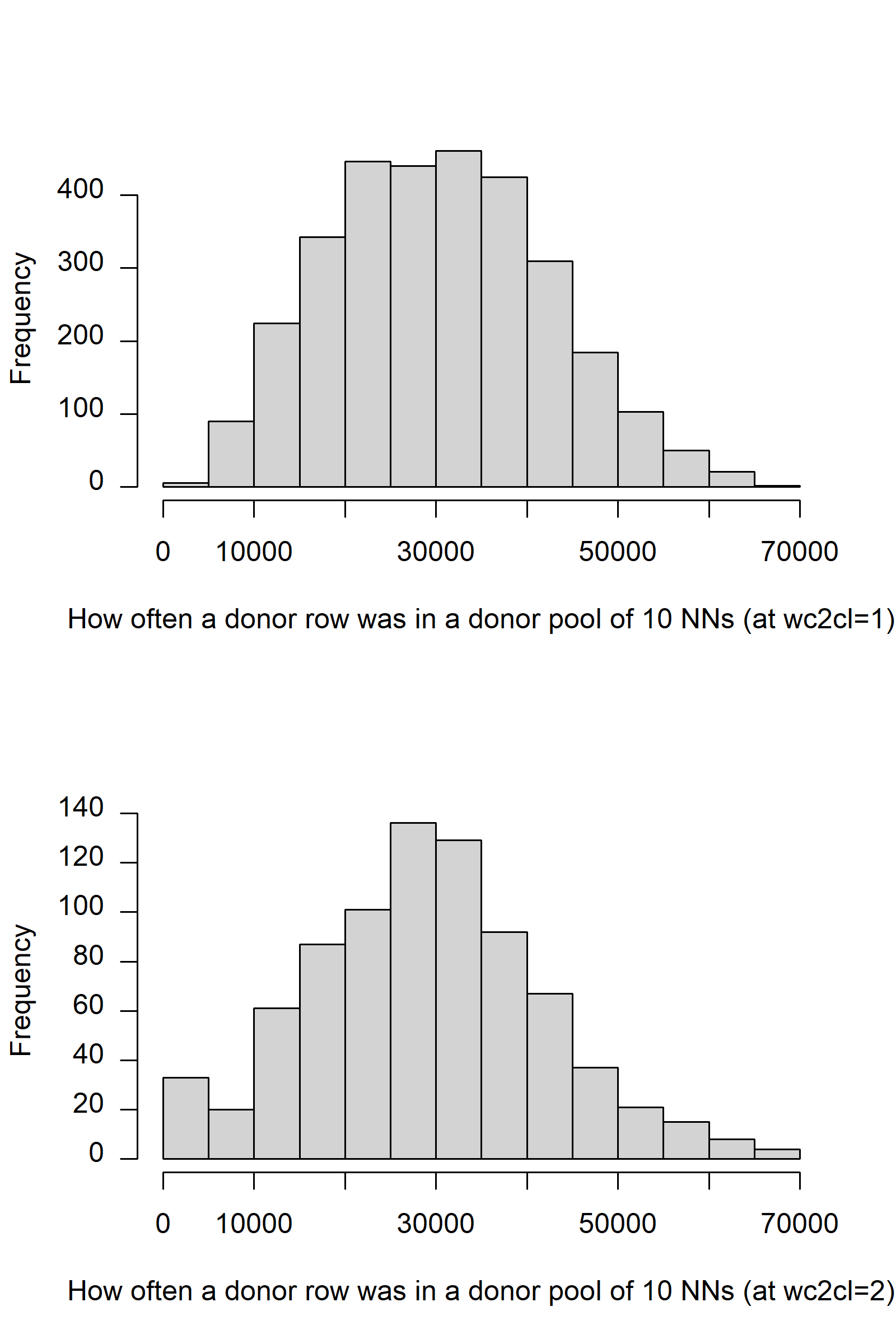}
    \includegraphics[width=.45\textwidth]{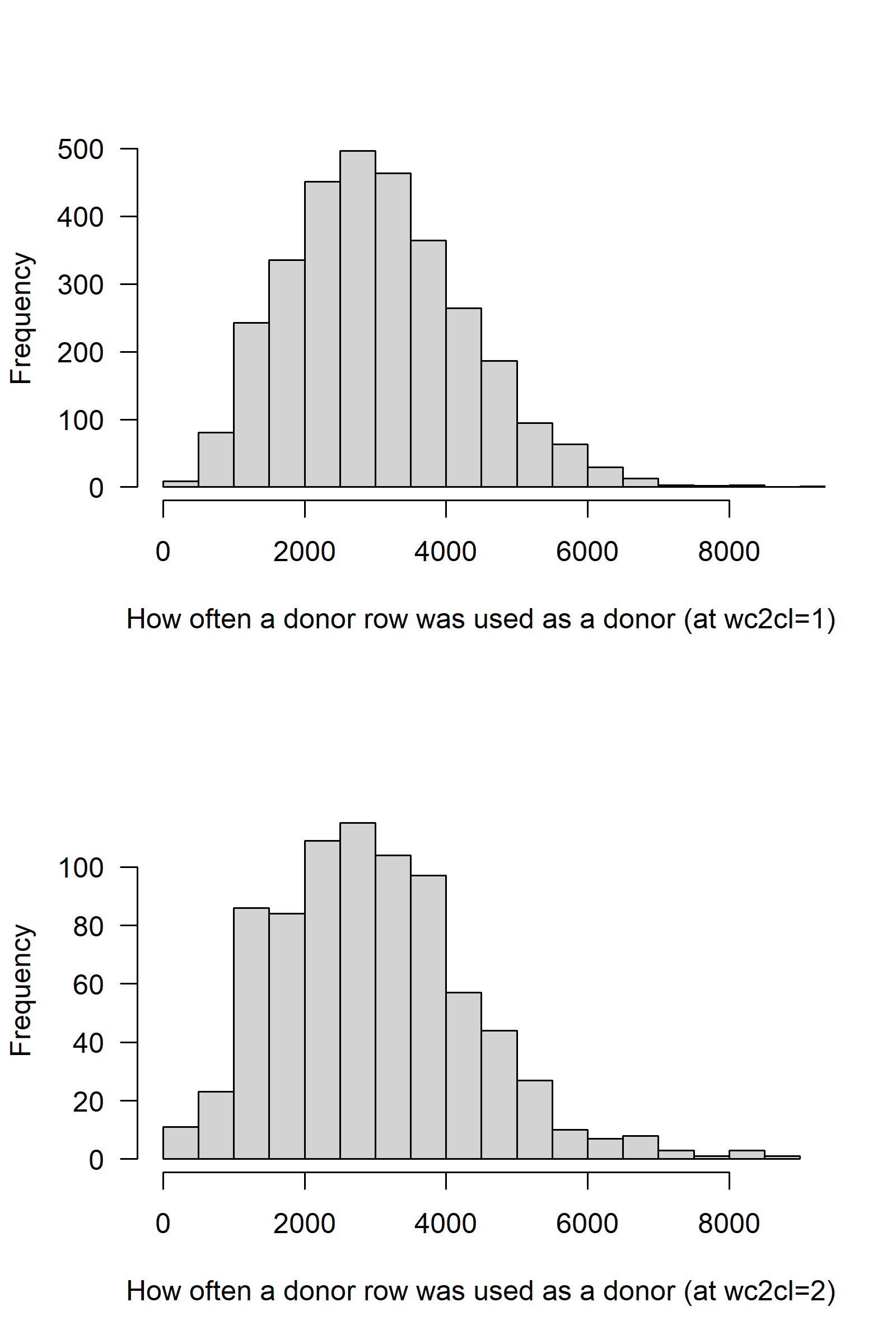}
    \caption{Count of how often each donor row was included in a donor pool of some recipient's 10 nearest neighbors (left) or actually used as a donor (right), for \texttt{wc2cl} classes 1 (tree, top) and 2 (non-tree, bottom).}
    \label{fig:DonorHists}
\end{figure}

Next, we checked how often each of the 3946 donor rows were used (Figure~\ref{fig:DonorHists}).  Recall that for each of the roughly 12 million recipient rows, KBAABB finds its 10 NNs among the donor rows and then uses weighted sampling to choose one donor. Hence, first we checked how often each donor was included in a donor pool of some recipient's 10 NNs (left side of Figure). For both \texttt{wc2cl} classes, most donors were chosen to be in 10,000 to 50,000 donor pools, though some were chosen for as many as 69,302 donor pools while others for as few as 1 or 0 donor pools. Every one of the 3101 donors in class 1 (tree) was in a donor pool at least once, and 811 of the 845 donors in class 2 (non-tree) were in a donor pool at least once. Next, we repeated the analysis but to see how often each donor row was actually used as a donor (right side of Figure). For both classes, most donors were actually used for 1,000 to 5,000 recipients, though some were used as many as 9,173 times while others were only used 1 or 0 times. Again every one of the 3101 donors in class 1 was used as a donor at least once, but 790 of the 845 donors in class 2 were used as a donor at least once. In short, only 55 of the possible donors were never used (all in \texttt{wc2cl} class 2); 34 of them were never in a donor pool, and the other 11 were in a pool but never happened to be chosen by the weighted sampling.
We are not concerned about these few never-used donors, because they all came from the non-tree class, in which most of the Y-variables are all 0s. 
The other possible donors were typically used a moderate number of times, with few over-used or under-used donors.

\begin{figure}
    \includegraphics[width=\textwidth]{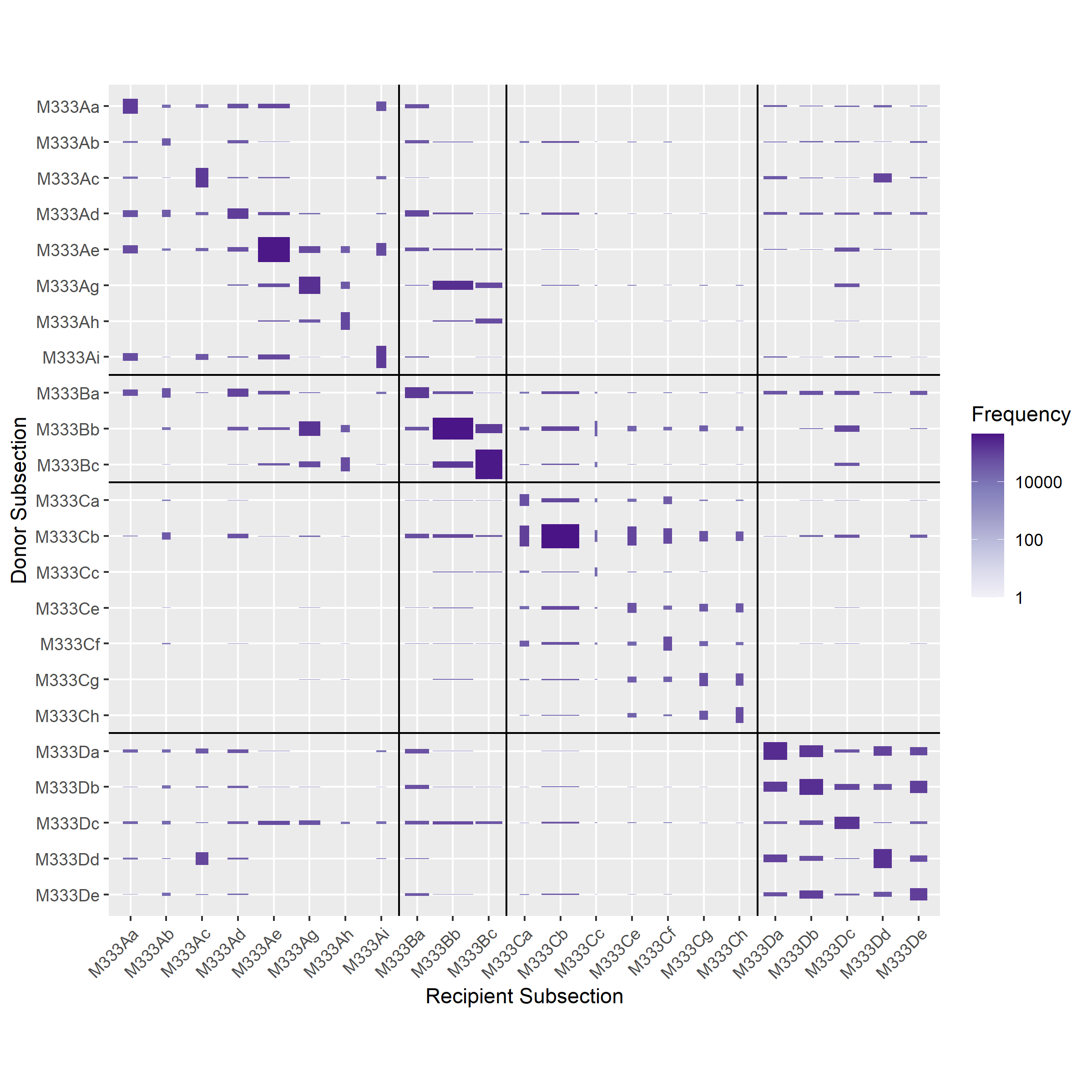}
    \caption{Count of how often donors from each subsection were used by recipients in each subsection.
    Box widths (fixed within each column) are proportional to the total number of pixels in each recipient subsection. Box heights (varying within each column) are proportional to the number of times a donor from that subsection was used for a recipient in that subsection, scaled so that boxes along the main diagonal would be squares if each conditional proportion (how often did that recipient use donors from its own subsection?) was equal to the overall proportion (31\% of recipient pixels got donors from their own subsection).
    }
    \label{fig:DonorRecipHeatmap}
\end{figure}

Recall that our 23 domains are ``subsections,'' nested within 4 ``sections,'' ecologically defined so that pixels should tend to be more similar within sections than across them.
We cross-tabulated the domain of each donor vs.\ each recipient to see whether recipients tended to get donors from the same subsection; from a different subsection but still the same section; or from elsewhere (Figure~\ref{fig:DonorRecipHeatmap}). Within each column (recipient), the highest frequencies tend to be in the corresponding row (donor) or another row in the same section (regions separated by black lines), showing that recipients often but not always tended to get donors from the same subsection or at least the same section. If all counts had been only on the main diagonal (each domain's imputed values all came from the same domain), we would have been concerned that the artificial population is too similar to simply taking copies of the survey sample. On the other hand, if counts were completely uniform, we would have been concerned that recipients are getting donors which do not actually resemble them. As it is, we feel comfortable seeing that our results fall between these two extremes.

Outside of these diagnostic checks, we also qualitatively assessed the spatial distribution of the artificial population generated in KBAABB compared to unweighted kNN approaches. For a discussion of this analysis, please see Supplementary material~\ref*{sec:Sensitivity}.

\subsection{Case study: using KBAABB to evaluate SAE methods in our FIA setting}\label{sec:CaseStudies}

The artificial population generated with KBAABB allows for an extensive empirical evaluation of SAE methods and strategies that may be implemented by researchers and others interested in the performance of small area estimators under realistic conditions for the population and sampling design of interest. In order to evaluate such small area estimators for average basal area per acre for M333, we created 2500 samples from the artificial population, where we sampled one pixel per hexagon for each hexagon that overlaps M333, as described in Supplementary material~\ref*{sec:Sampling}. We assess the small area estimators described in Section~\ref{sec:CaseStudiesMethods} based on their estimates and MSE estimates of average basal area per acre within each subsection in M333. 

\begin{figure}
    \includegraphics[width=\textwidth]{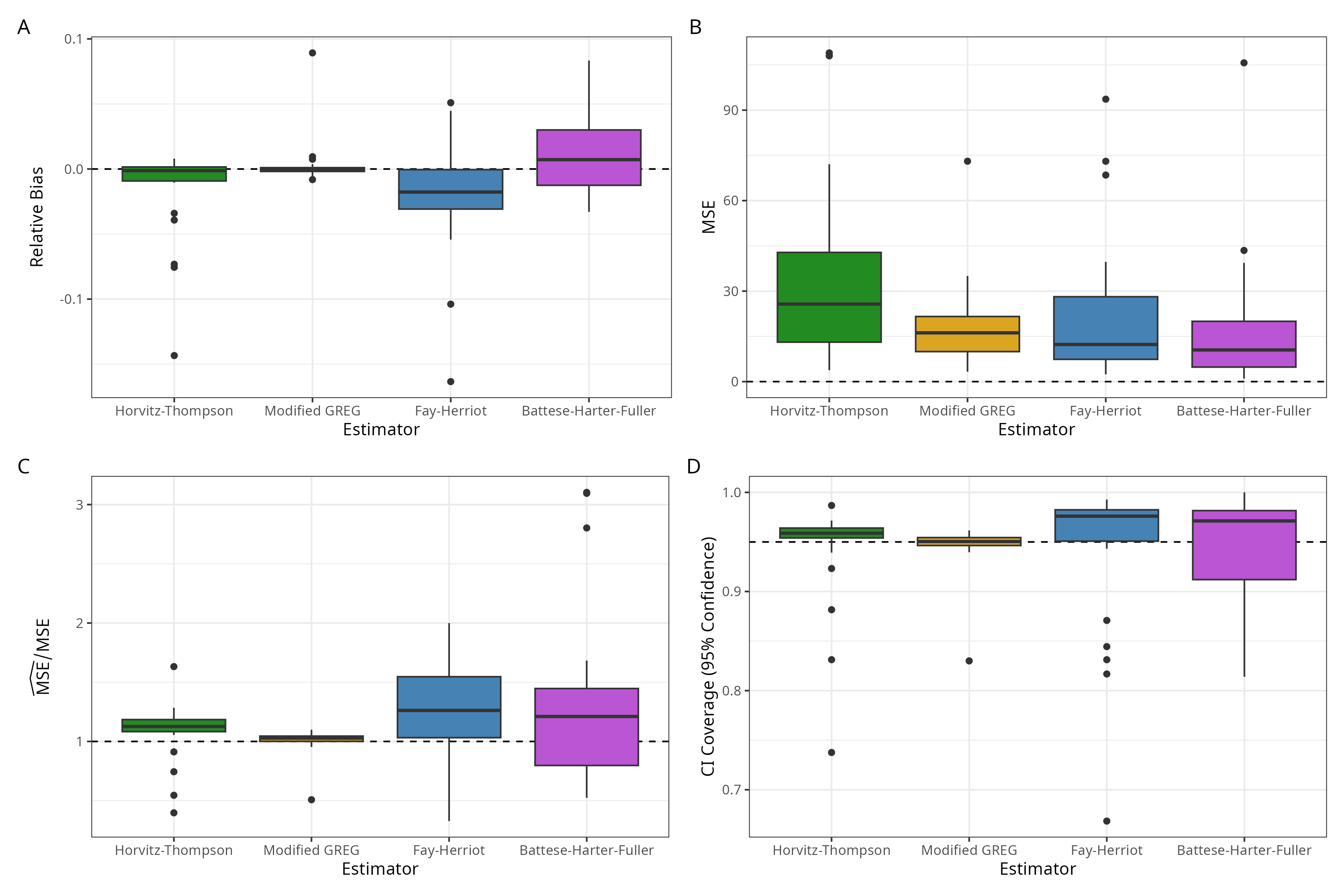}
    \caption{Performance metrics of each estimator for each subsection displayed as side-by-side boxplots. Green represents the Horvitz-Thompson, yellow the modified GREG, blue the area-level EBLUP estimator based on the Fay-Herriot, and purple the unit-level EBLUP estimator based on the Battese-Harter-Fuller. Subfigure A displays the relative bias metric, subfigure B displays the empirical MSE, subfigure C displays the ratio of estimated and empirical MSE, and subfigure D displays the 95\% confidence interval coverage rate. For subfigures A-D, the horizontal dashed line has an intercept of 0, 0, 1, and 0.95, respectively, representing the ideal value for each metric.}
    \label{fig:GraysonMetrics}
\end{figure}
\begin{figure}
    \includegraphics[width=\textwidth]{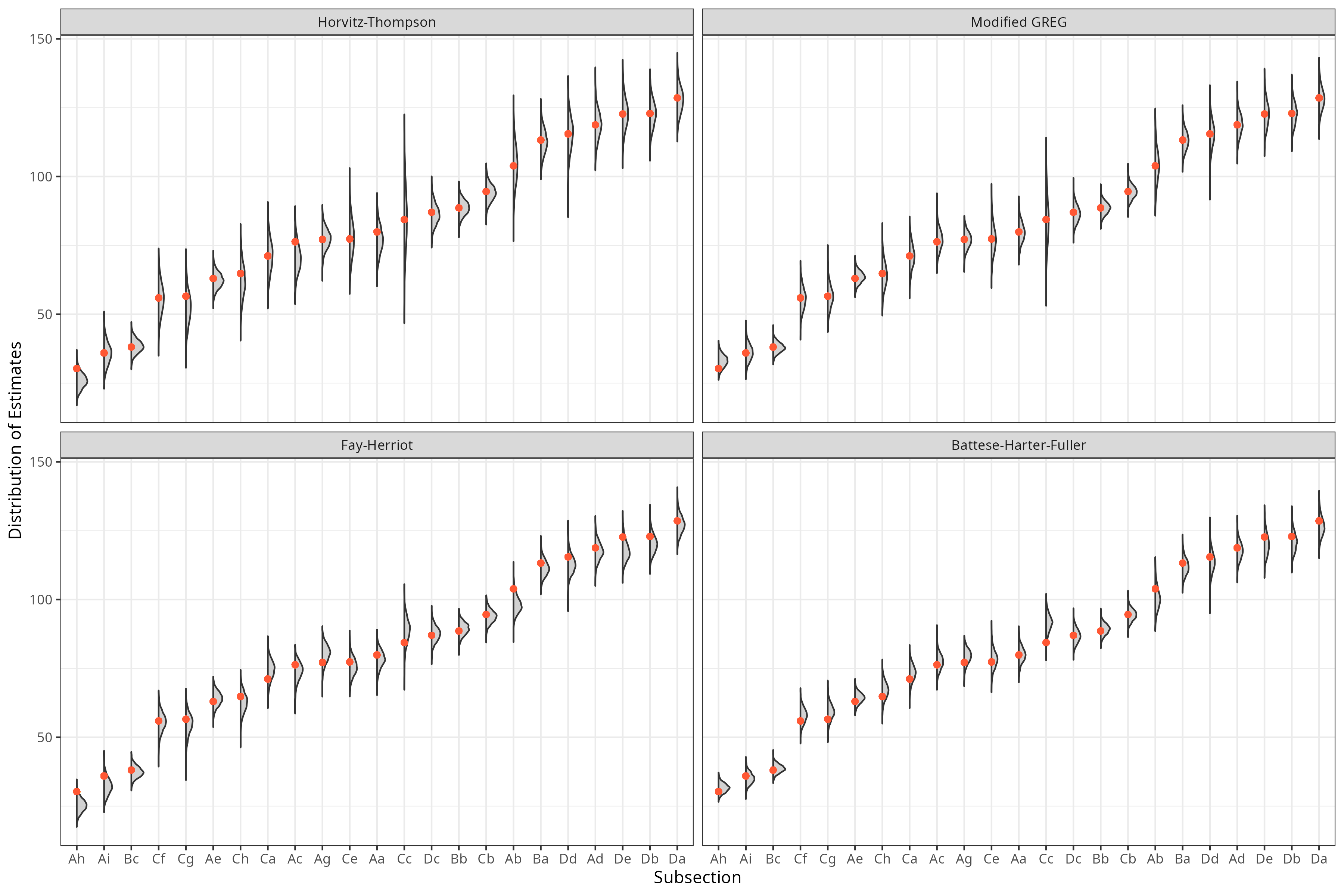}
    \caption{Distributions of estimates produced for each estimator and subsection, along with true KBAABB population parameter value (red dot). The subsections are mapped to the x-axis, and the distribution estimates and true parameter values are mapped to the y-axis. Each panel corresponds to a different estimator. Subsections are ordered by true average basal area per acre from the KBAABB population, least to greatest.}
    \label{fig:GraysonDists}
\end{figure}

In order to assess the metrics described in Section~\ref{sec:CaseStudiesMethods}, we turn to Figure~\ref{fig:GraysonMetrics}. Each subfigure displays a metric for each subsection and estimator combination; in particular subfigure A displays the relative bias, subfigure B displays the MSE, subfigure C displays the MSE ratio, and subfigure D displays the confidence interval coverage. For the two design-based estimators, the relative bias remains low, as we would expect. However, for the model-based estimators we see a larger spread of values for relative bias. On median, the area- and unit-level EBLUP estimators exhibit negative and positive relative bias, respectively. Further, the distribution of the area-level EBLUP's bias is primarily negative, with its interquartile range entirely following below the 0 on the y-axis. While examining the empirical relative bias can be informative, it can also be useful to look at the distribution of estimates produced over each simulation rep in order to get a fuller sense of the estimator's performance across samples. This is shown in Figure~\ref{fig:GraysonDists}. From Figure~\ref{fig:GraysonMetrics} alone we might be concerned with some of the biases that the Horvitz-Thompson estimator is exhibiting, but upon examination in Figure~\ref{fig:GraysonDists} we can see how variable the estimator is, and that it does not appear to be producing concerning bias in any subsection. Contrastingly, the model-based estimators display some more concerning characteristics: in subsections such as M333Ah (Okanogan Semi-Arid Foothills) and M333Cc (Mission-Swan Valley-Flathead River), we see the model-based estimators producing precise estimates that are consistently over- or under-estimating the parameter of interest, leading to bias in the estimator. This bias may be attributed to a variety of causes, but upon ecological inspection of the subsections, we see that, compared to the rest of the M333 province, these are lower elevation regions which likely exhibit very different qualities than the otherwise mountainous majority of M333, making it difficult for the model's random effect to pool accurately. The modified GREG estimator is both moderately precise and exhibits less concerning distributions than the model-based approaches, although the modified GREG does still have trouble capturing the true mean in M333Ah. 

Returning to Figure~\ref{fig:GraysonMetrics}, we see in subfigure B the MSE of model-based and -assisted estimators are competitive in this case study, with the two model-based approaches exhibiting the lowest MSE on median, but the modified GREG estimator displaying similar values. Subfigure C displays the MSE ratio, i.e. how well an estimator's MSE is estimated, and we see here the Horvitz-Thompson and modified GREG estimators with values right around 1, and the model-based estimators show an overestimation of MSE, on median. Further, in subfigure D we see the 95\% confidence interval coverage looks very sensible for the Horvitz-Thompson and modified GREG estimators, and slightly more instances of over- or under-coverage for the model-based estimators. This makes sense in light of subfigure C where we see an over-estimation of the uncertainty in these estimators.

\begin{figure}
    \includegraphics[width=\textwidth]{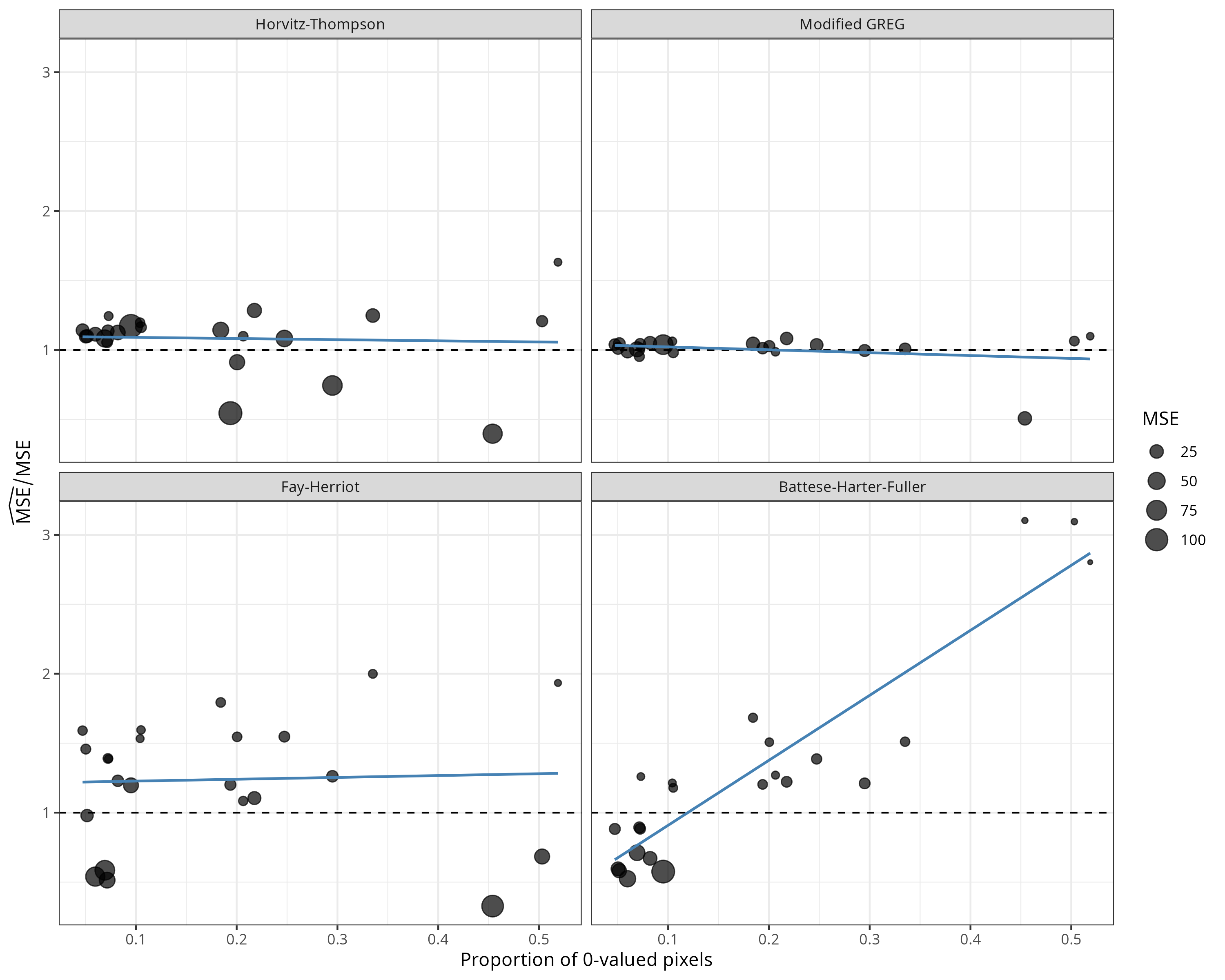}
    \caption{Ratio of estimated MSE and empirical MSE across every domain and estimator by proportion of zero-valued pixels for the response variable in the population, sized by empirical MSE. The MSE ratio is mapped to the y-axis, and the proportion of zero-valued pixels is mapped to the x-axis. Empirical MSE is mapped to point size. The dashed horizontal line has a slope of 0 and intercept of 1, and represents perfect MSE estimation. Blue lines are least-squares regression lines.}
    \label{fig:GraysonMSE}
\end{figure}

In light of Figure~\ref{fig:GraysonMetrics}, subfigure C, we wanted to understand if the poor MSE estimation of some estimators can be attributed to any systematic properties in the areas of interest, and to assess this we turn to Figure~\ref{fig:GraysonMSE}, which displays the MSE ratio by proportion of zero-valued pixels in the population for the response variable, with points sized by MSE. The Horvitz-Thompson, modified GREG, and area-level EBLUP estimator do not exhibit a relationship between the displayed variables; however, we see that the area-level EBLUP does over-estimate the MSE for subsections with small MSEs, and under-estimates MSE for subsections with large MSEs. This is somewhat expected with the pooling that occurs when fitting a mixed model. We also see this trend for the unit-level EBLUP estimator. Strikingly though, the unit-level EBLUP estimator's MSE ratio has a strong positive linear relationship with the proportion of zero-valued pixels in the population for the response variable. We see that when the proportion of zero-valued pixels is greater than approximately 0.1, the MSEs tend to be smaller but are also over-estimated. This is likely due to model-misspecification that can occur at the unit-level due to zero-inflation in the data. The forest inventory data used here is positive and continuous sans zeros, and we expect as the proportion of zeros in a subsection increases, that the model will be increasingly misspecified. Zero-inflated estimators have been proposed to deal with the model-misspecification discussed here \citep{finley2011hierarchical, white2024small}, but this systematic issue in estimating the MSE of small area estimates is not discussed. In order to obtain the precision of a unit-level, model-based approach without the perils of the unit-level EBLUP shown here, it may be worthwhile to consider the approaches taken by \citet{finley2011hierarchical} or \citet{white2024small}.

From this case study, we are able to obtain much information about the estimators considered for average basal area per acre in M333. First of all, the Horvitz-Thompson estimator appears to be performing as we expect, which is a good check that our simulation is producing sensible results. Next, the modified GREG and area-level EBLUP both appear to be decent estimators for these data and areas of interest. Both estimators produce sensible estimates; however, in some outlying subsections we do see estimator bias appear. Finally, the unit-level EBLUP displays concerning results regarding poor and systematic over-estimation of the MSE in this setting. While the unit-level EBLUP was the most precise estimator in terms of MSE (on median, Figure~\ref{fig:GraysonMetrics}), its MSE estimation was unreliable. Given the choice between the four estimators,
the modified GREG has been empirically shown to have the best balance of low bias, low MSEs, accurate MSE estimation, and correct confidence interval coverage on this artificial population.

\section{Discussion}\label{sec:discussion}

The need to properly assess small area estimators is a high priority for those making inferences and management decisions with forest inventory data. SAE methods promise precise estimates in constrained geographic and temporal regions, but merely fitting a small area model is insufficient to making well-informed inferences. We presented KBAABB to help empirically assess SAE methods. We find KBAABB to be well-suited for assessing estimators on forest inventory data in our case study, but KBAABB is suitable more generally when we have rich and complete auxiliary data for the entire population of interest. Of course, not all surveys have unit-level auxiliary data for the entire population of interest;
but when they do, we argue that KBAABB's artificial populations are realistic, fine-grained, and not biased in favor of any particular SAE model, making KBAABB appropriate for ``satisfactory extrapolation'' from simulation studies to SAE estimator performance for the case at hand.

One direction for future work is nonresponse. FIA surveys do experience some nonresponse (for instance, land owners may not admit a survey field crew, or weather conditions may make it impossible to visit a survey plot on schedule), though less than a typical demographic survey and this nonresponse is usually treated as missing at random (MAR) within a given population \citep{bechtold2005enhanced}.
So far, in generating and sampling from our artificial populations, we have assumed that nonresponse was missing completely at random (MCAR) and ignorable.
However, in demographic applications we typically take extra care to ensure that imputation is used appropriately given the posited nonresponse mechanism, which is often MAR or missing not at random (MNAR) rather than MCAR.
It should be possible to add a simulated non-response stage after KBAABB's sampling stage, to mimic any non-response behavior seen in the real survey.
Appropriate use of KBAABB for generating artificial populations when the survey data had non-MCAR nonresponse is an open question.

Another future direction is the interplay between variable selection for KBAABB matching vs.\ for SAE modeling. KBAABB is not biased towards any of the common parametric SAE model forms. However, it remains to be seen whether the choice of variables for KBAABB matching may bias the performance of SAE models, when comparing models whose explanatory variables do vs.\ do not include the matching variables. For the present paper, we side-step this issue by using subject-matter expertise to choose the KBAABB matching variables, and by only using a subset of those variables in the SAE models in our case studies.

We have also not touched on a few complications specific to the FIA data: subplots and condition classes, data vintage, or spatial smoothing. Briefly, condition classes and subplots are more fine-grained levels of detail about the survey data collected at each plot. Our artificial population has been generated only at the plot level so far, and we may extend our methods to work at these other, smaller levels. Meanwhile, the issue of data vintage refers to accounting for the date at which each variable was measured. The survey data comes from a panel design, in which different plots are sampled in different years; and the surveyed pixels' $X$ variables may not be from the same year as the corresponding $Y$. Although most of the $X$s change very slowly if at all from year to year (such as elevation or precipitation), others might be more timely (such as tcc, wc2cl, or tnt which should change as forests grow, become harvested, experience wildfires, etc.). Both our artificial population and our SAE models might benefit from a closer match between the vintages of $X$ and of $Y$, though prior work by FIA researchers suggests that common SAE models are fairly robust to data vintage in this population.
Finally, in order to assess estimators with explicit spatial components, future work will incorporate spatial smoothing to the imputed $Y$ values after running KBAABB.

Aside from such details that are specific to our FIA data sources, we believe that KBAABB is a useful approach that could be widely applied to generate artificial populations and evaluate small area estimators for any survey where rich, unit-level auxiliary data is available.

\section*{Software} We are developing an R package \texttt{kbaabb} which provides tools for generating an artificial population via KBAABB. The package is available at
\url{https://github.com/graysonwhite/kbaabb/}. While we cannot share the proprietary FIA data used in this paper, the R package includes a worked example based on public data.

We have also created a draft R Shiny application \citep{chang2022shiny} for exploring the case study's artificial population and several estimators. The app can be accessed at \url{https://civilstat.shinyapps.io/fia-simpop-app/}
and the code is available at \url{https://github.com/ColbyStatSvyRsch/FIA-simpop-app}. For details and screenshots, see Supplementary material~\ref*{sec:Shiny}.

\section*{Funding} 
The authors disclosed receipt of the following financial support for the research, authorship, and/or publication of this article: This work was supported by the National Council for Air and Stream Improvement [grant number FO-SFG-2673 to J.A.W.].

\section*{Supplementary material}

The following supplementary material is available at \textit{Forestry} online.

\textbf{S.1:} Sensitivity to $k$ and weighted sampling in our FIA setting

\textbf{S.2:} KBAABB implementation and sampling in our FIA setting

\textbf{S.2.1:} KBAABB implementation in our FIA setting

\textbf{S.2.2:} Sampling for design-based simulations in our FIA setting

\textbf{S.3:} Exploring KBAABB output with an R Shiny application

\section*{Conflict of interest statement}
None declared.

\section*{Data availability statement}

The data include confidential plot data, which can not be shared publicly. FIA data can be accessed through the FIA DataMart (\url{https://apps.fs.usda.gov/fia/datamart/datamart.html}). Requests for data used here or other requests including confidential data should be directed to FIA's Spatial Data Services (\url{https://www.fs.usda.gov/research/programs/fia/sds}).

\printbibliography

\newpage

\appendix
\begin{refsection}
\renewcommand{\thesection}{S.\arabic{section}}    

\resetlinenumber
\section*{{\huge Supplementary material}}
\setcounter{page}{1}
\setcounter{figure}{0}
\renewcommand{\thefigure}{S\arabic{figure}}

\section{Sensitivity to $k$ and weighted sampling in our FIA setting}\label{sec:Sensitivity}

Besides KBAABB, we also tried generating artificial populations by sampling uniformly from kNN with values of $k \in \{1,5,10,20,50,100\}$. We expected that $k$ of 10 or less might be reasonable, while much larger values would lead to artificial populations whose Y values were too close to being uniformly random, but we included larger $k$ values in order to check this. Indeed, results for KBAABB, $k$=1, and $k$=5 looked fairly similar, but $k$=10 or above were clearly less realistic. Among KBAABB, $k$=1, and $k$=5, our forestry subject-matter-experts believed that KBAABB looks more realistic than $k$=1 or $k$=5, and our statisticians believed that KBAABB was better justified than $k$=1.

\begin{figure}
    \includegraphics[width=\textwidth]{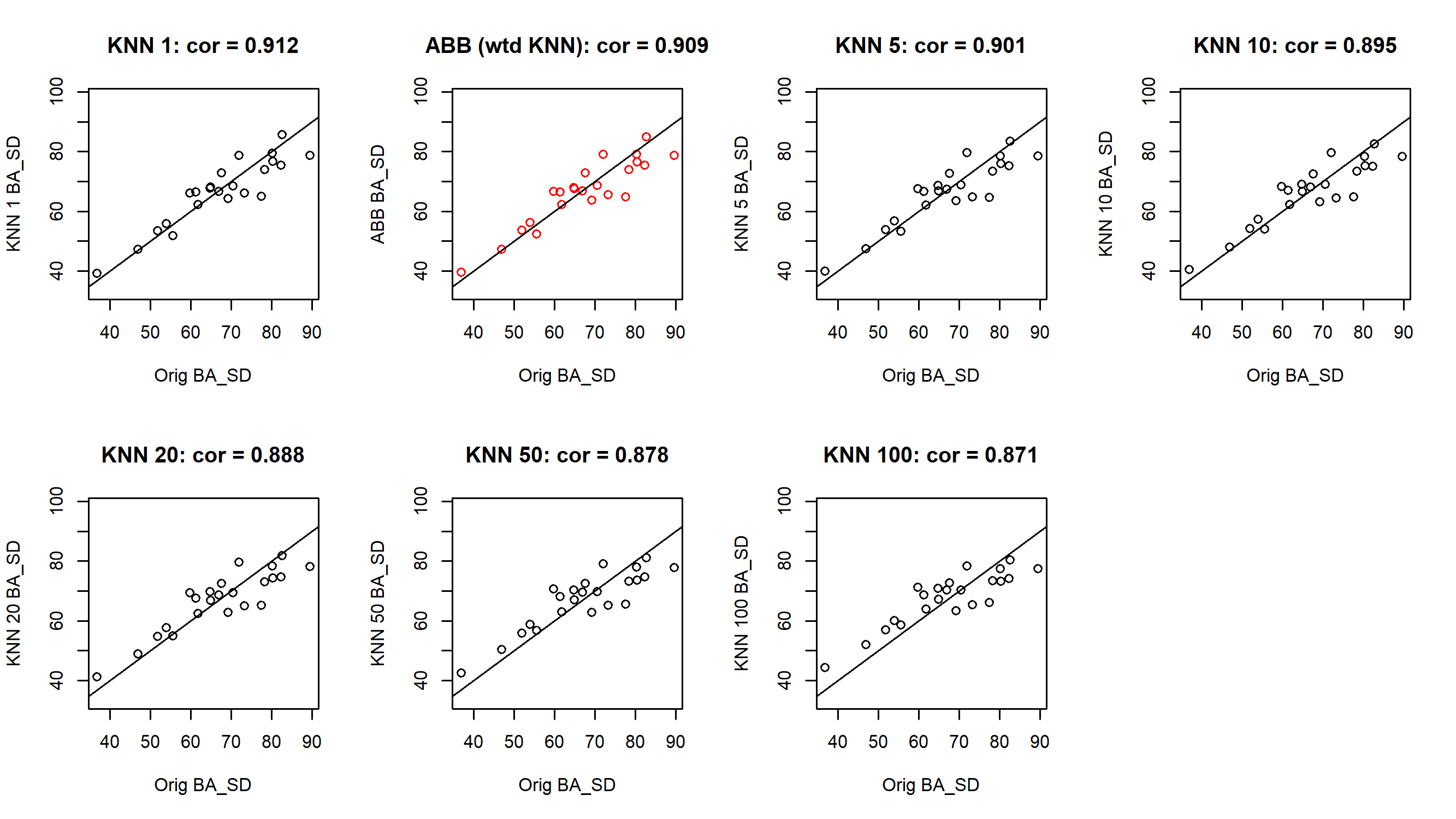}
    \caption{Original vs imputed SDs of average basal area per acre for each subsection, for KBAABB (red) and for kNN at several different $k$. All plots show $y=x$ line as a common reference.}
    \label{fig:SDs_BA}
\end{figure}

First we checked how $k$ affected the variability in imputed Y-values within each domain. Using average basal area per acre as an example, Figure~\ref{fig:SDs_BA} shows that for KBAABB as well as kNN with small values of $k$, each artificial population's domain-level SDs are strongly but not perfectly correlated with their original-sample counterparts. As $k$ grows, we see the correlations get weaker, because for large $k$ we are drawing uniformly from a large donor pool where the donors do not necessarily resemble the recipient. Indeed, for large enough $k$ (larger than any shown here), we would simply be sampling donors uniformly from the whole sample and the scatterplot would be essentially horizontal.

\begin{figure}
    \includegraphics[width=.9\textwidth]{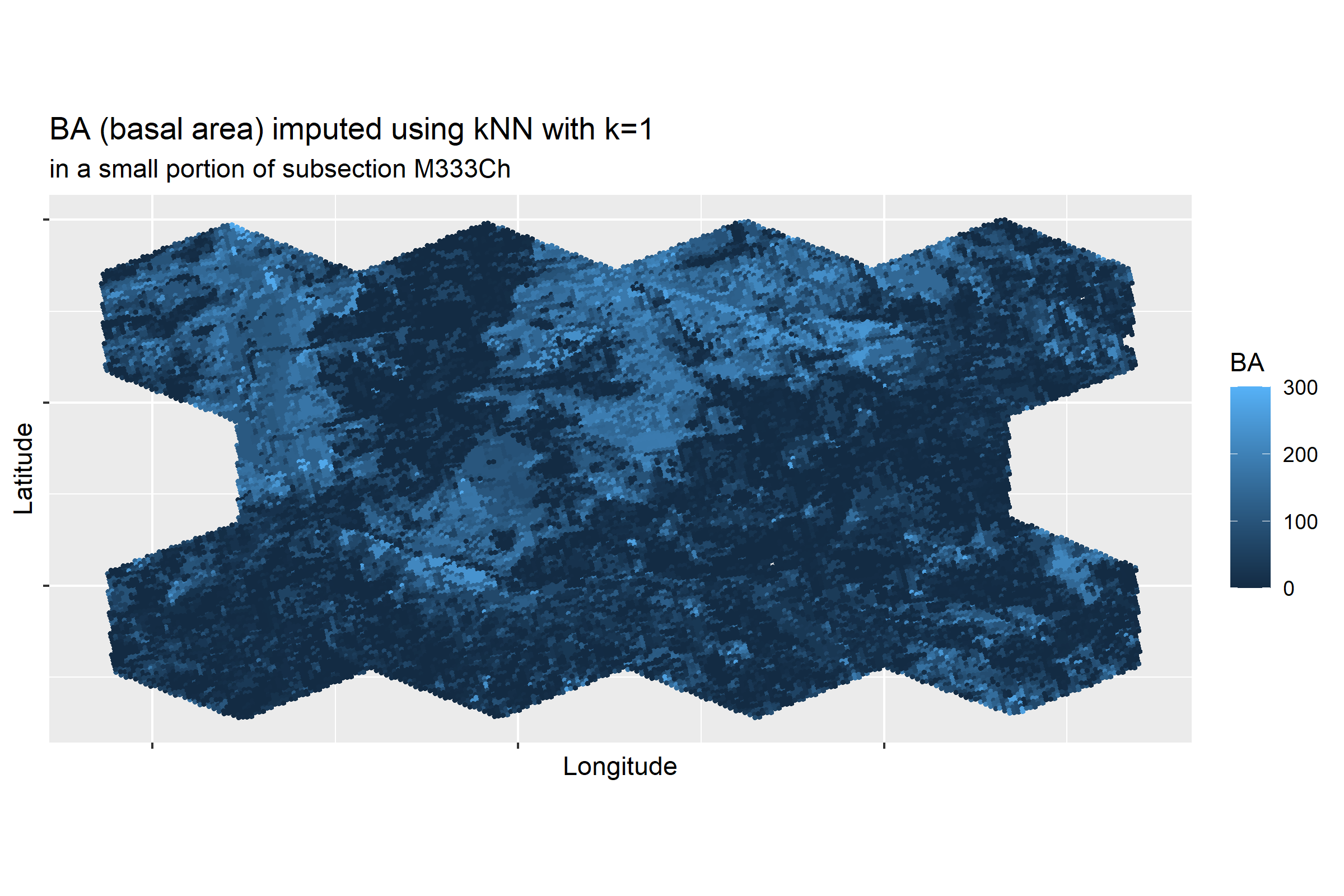}
    \includegraphics[width=.9\textwidth]{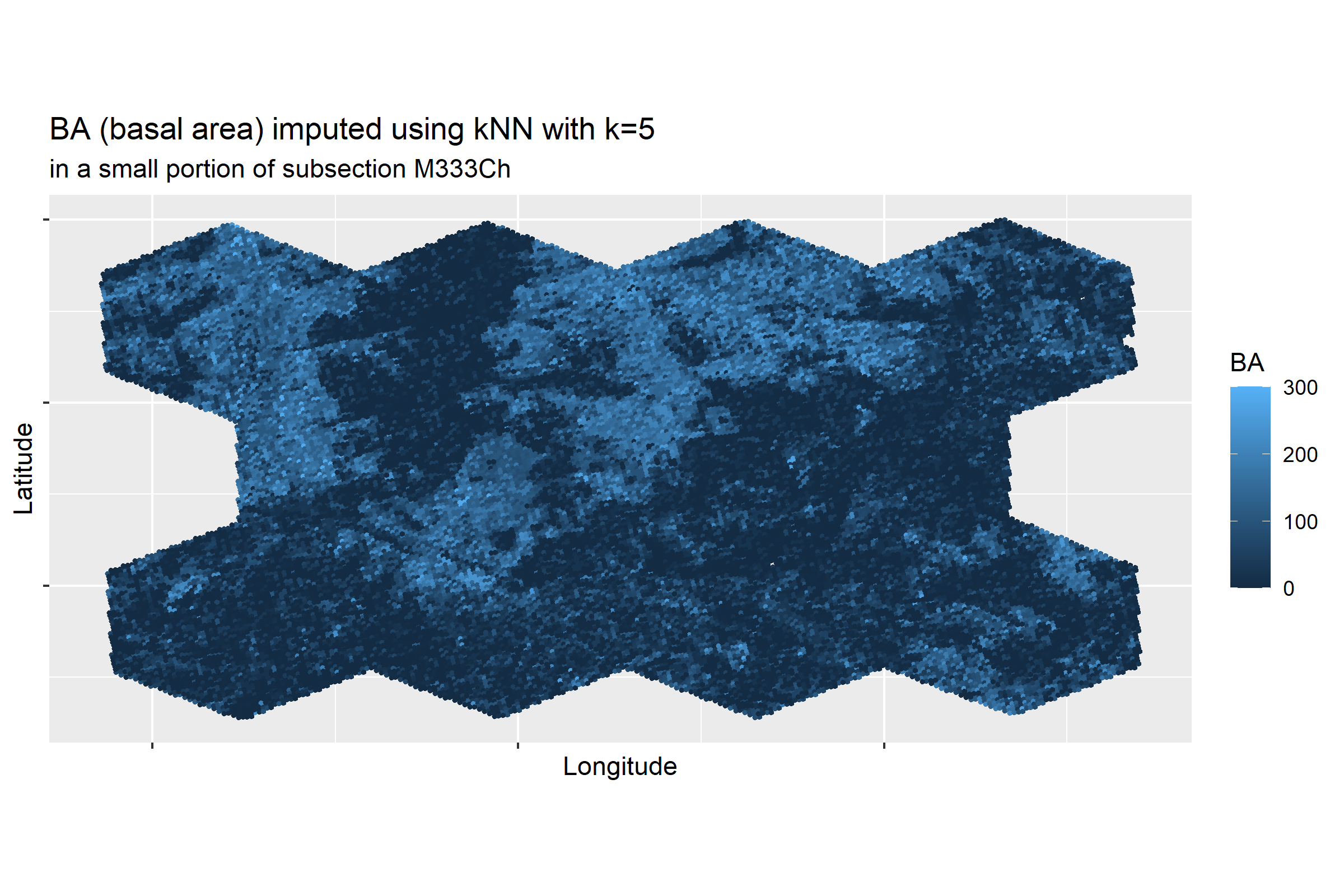}
    \caption{kNN with $k$=1 (above) and $k$=5 (below) imputations of average basal area per acre in a portion of subsection M333Ch.}
    \label{fig:Map_BA_k1and5}
\end{figure}

\begin{figure}
    \includegraphics[width=.9\textwidth]{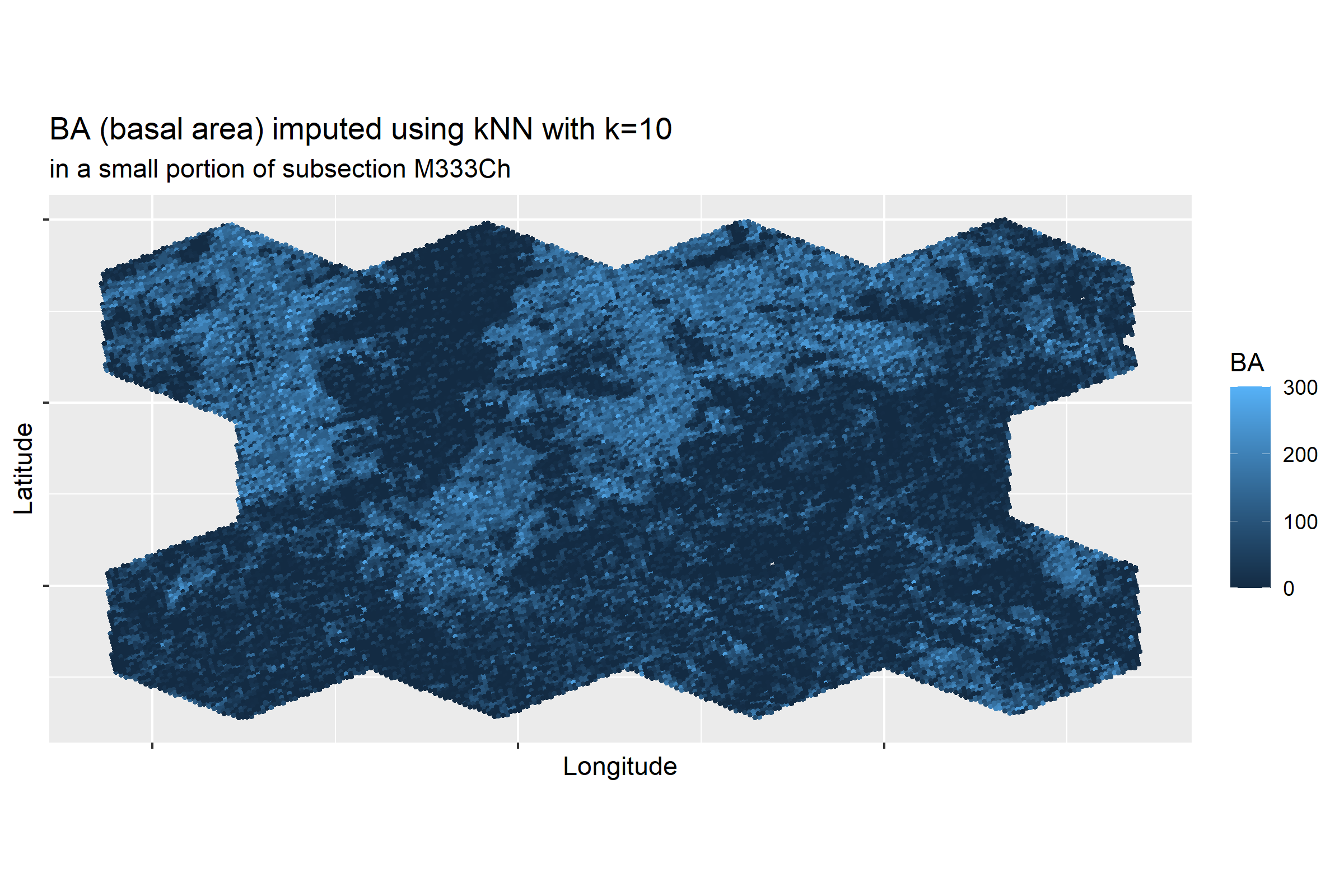}
    \includegraphics[width=.9\textwidth]{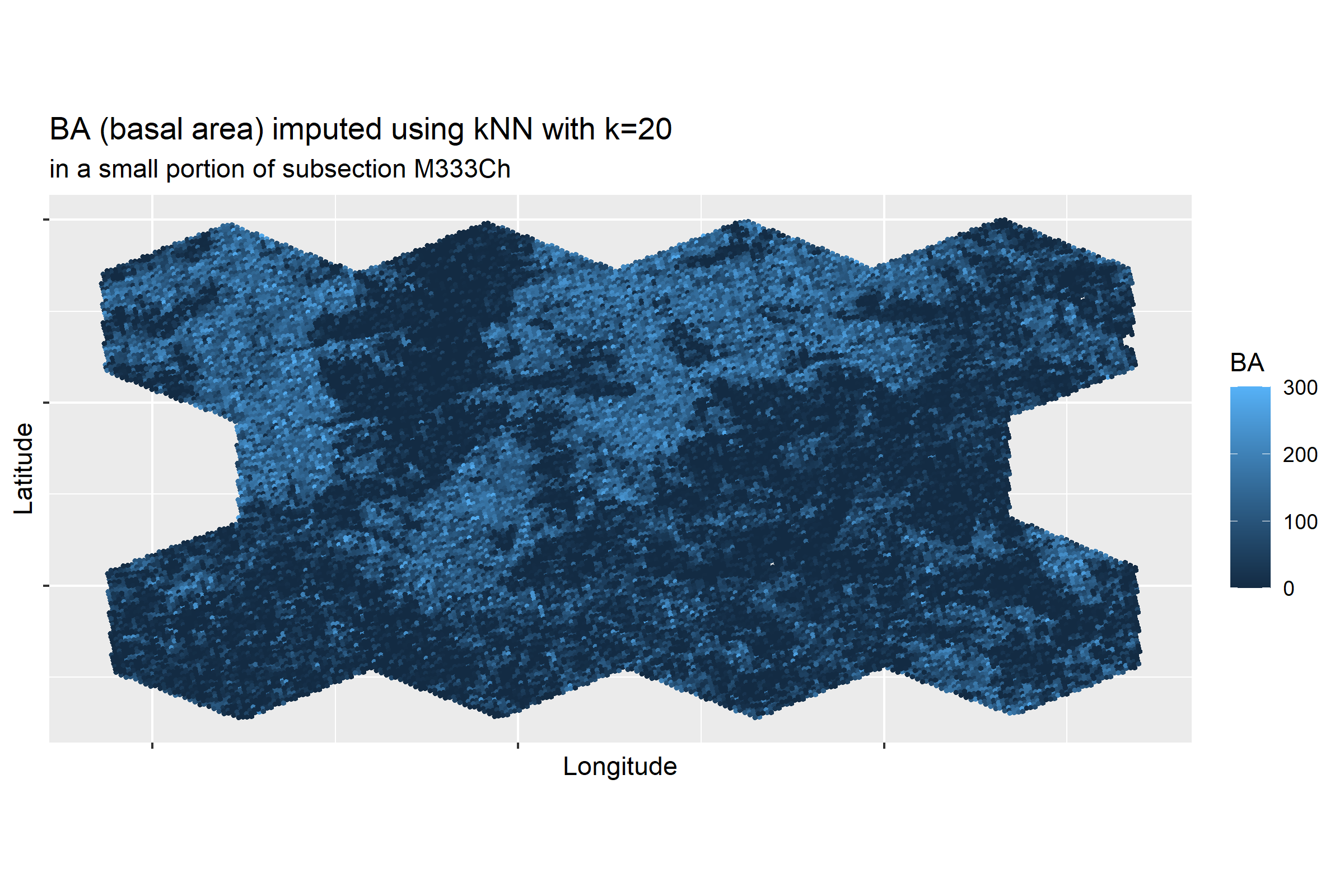}
    \caption{kNN with $k$=10 (above) and $k$=20 (below) imputations of average basal area per acre in a portion of subsection M333Ch.}
    \label{fig:Map_BA_k10and20}
\end{figure}

\begin{figure}
    \includegraphics[width=.9\textwidth]{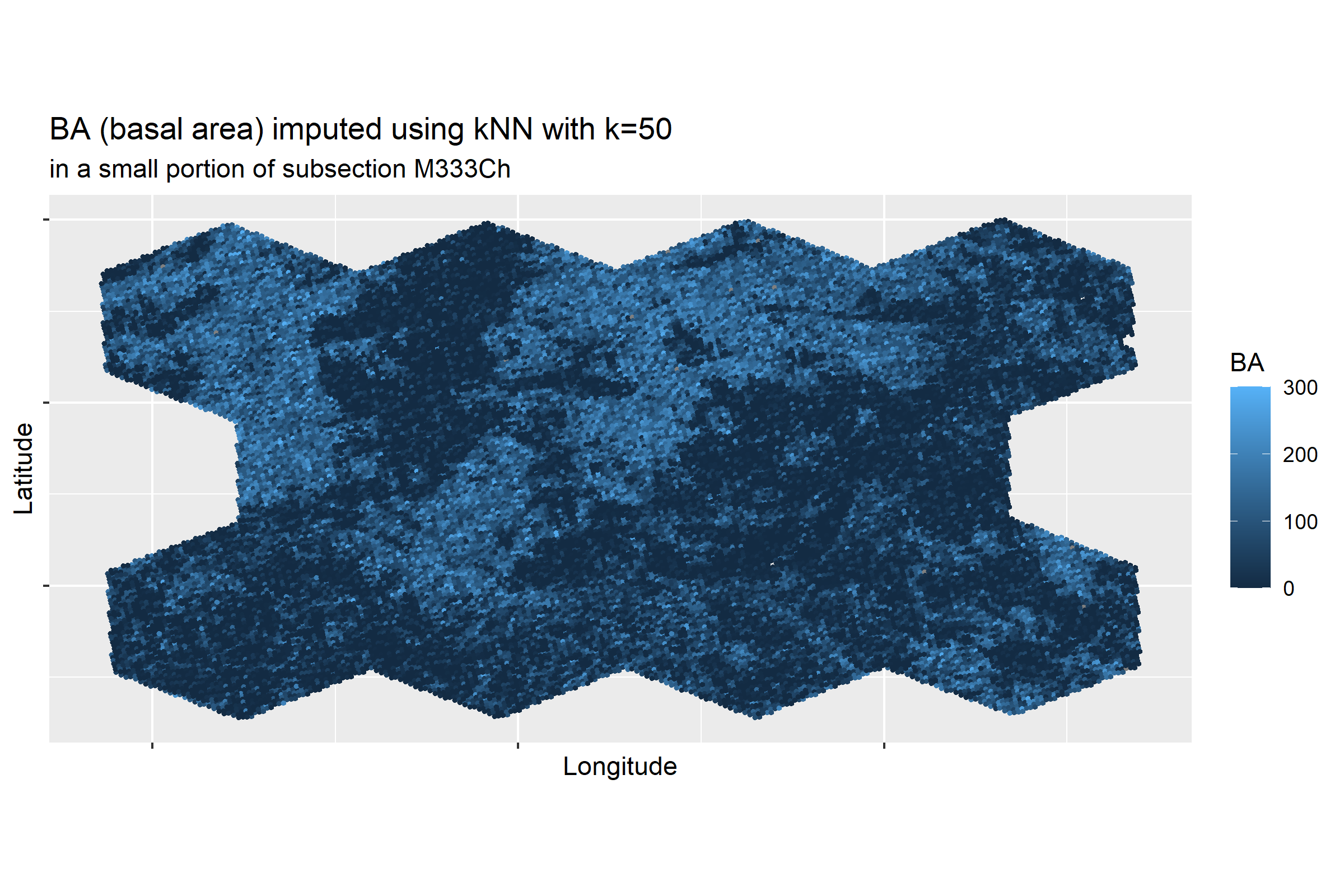}
    \includegraphics[width=.9\textwidth]{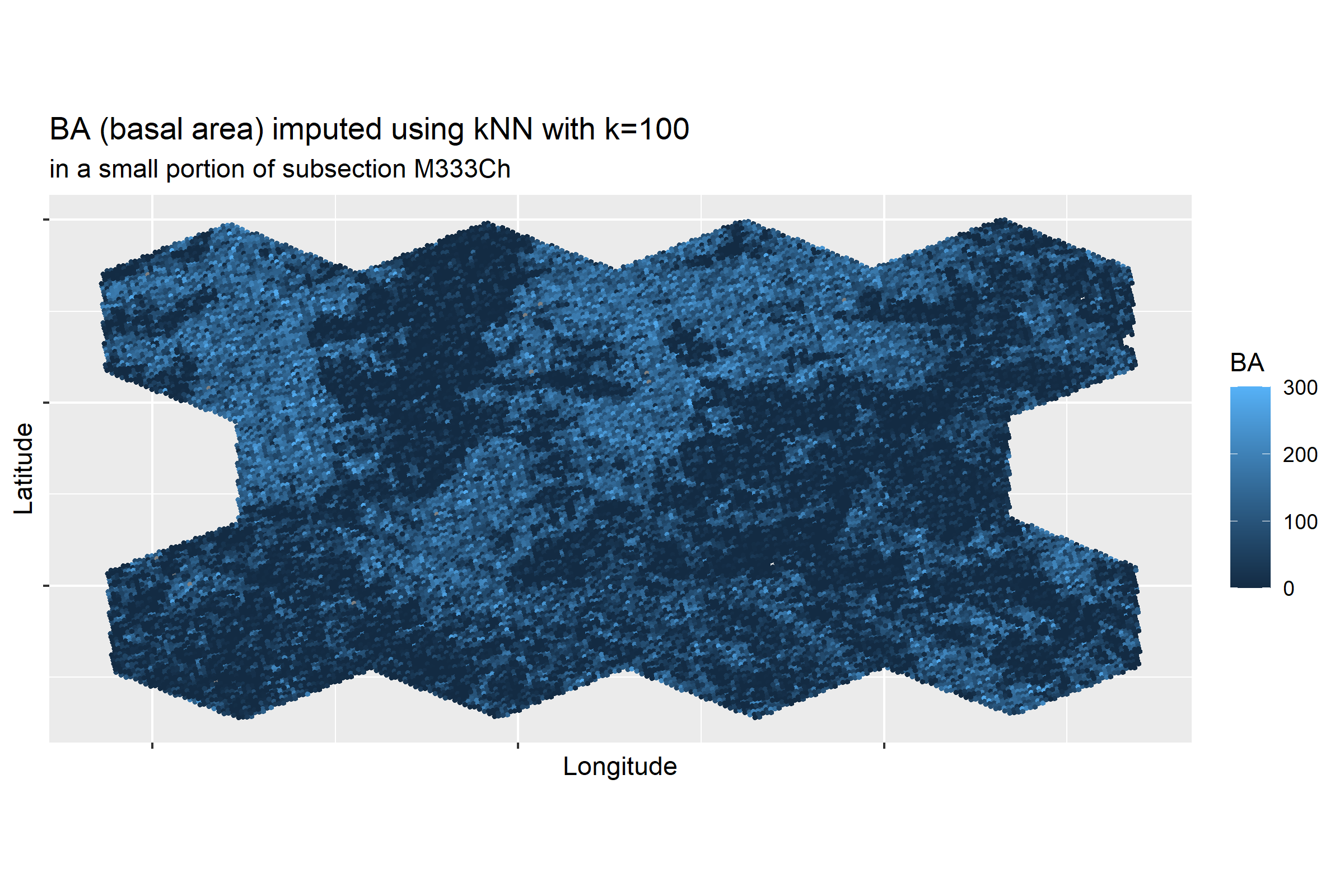}
    \caption{kNN with $k$=50 (above) and $k$=100 (below) imputations of average basal area per acre in a portion of subsection M333Ch.}
    \label{fig:Map_BA_k50and100}
\end{figure}

\begin{figure}
    \includegraphics[width=.9\textwidth]{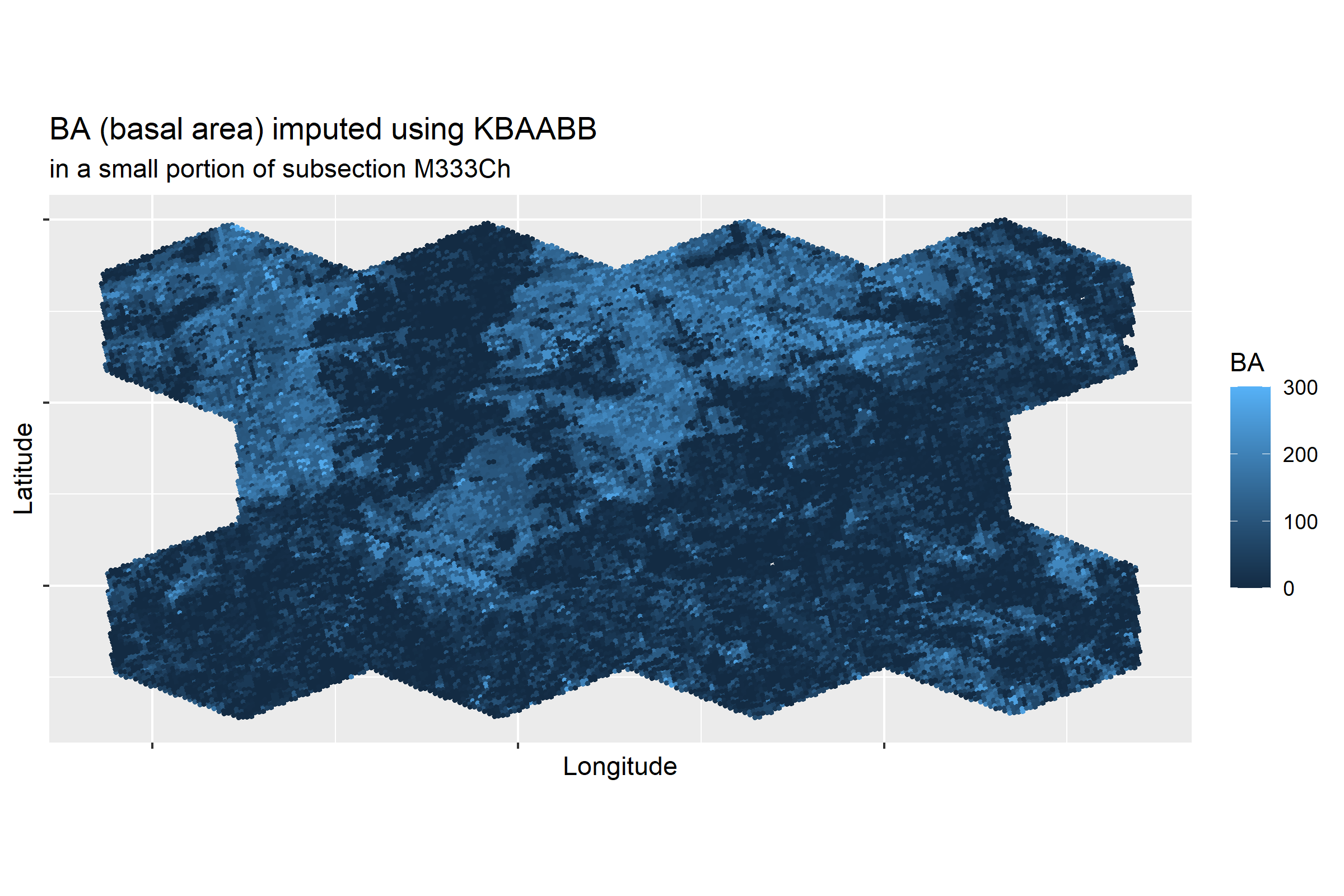}
    \caption{KBAABB imputation of average basal area per acre in a portion of subsection M333Ch.}
    \label{fig:Map_BA_KBAABB}
\end{figure}

\begin{figure}
    \includegraphics[width=.9\textwidth]{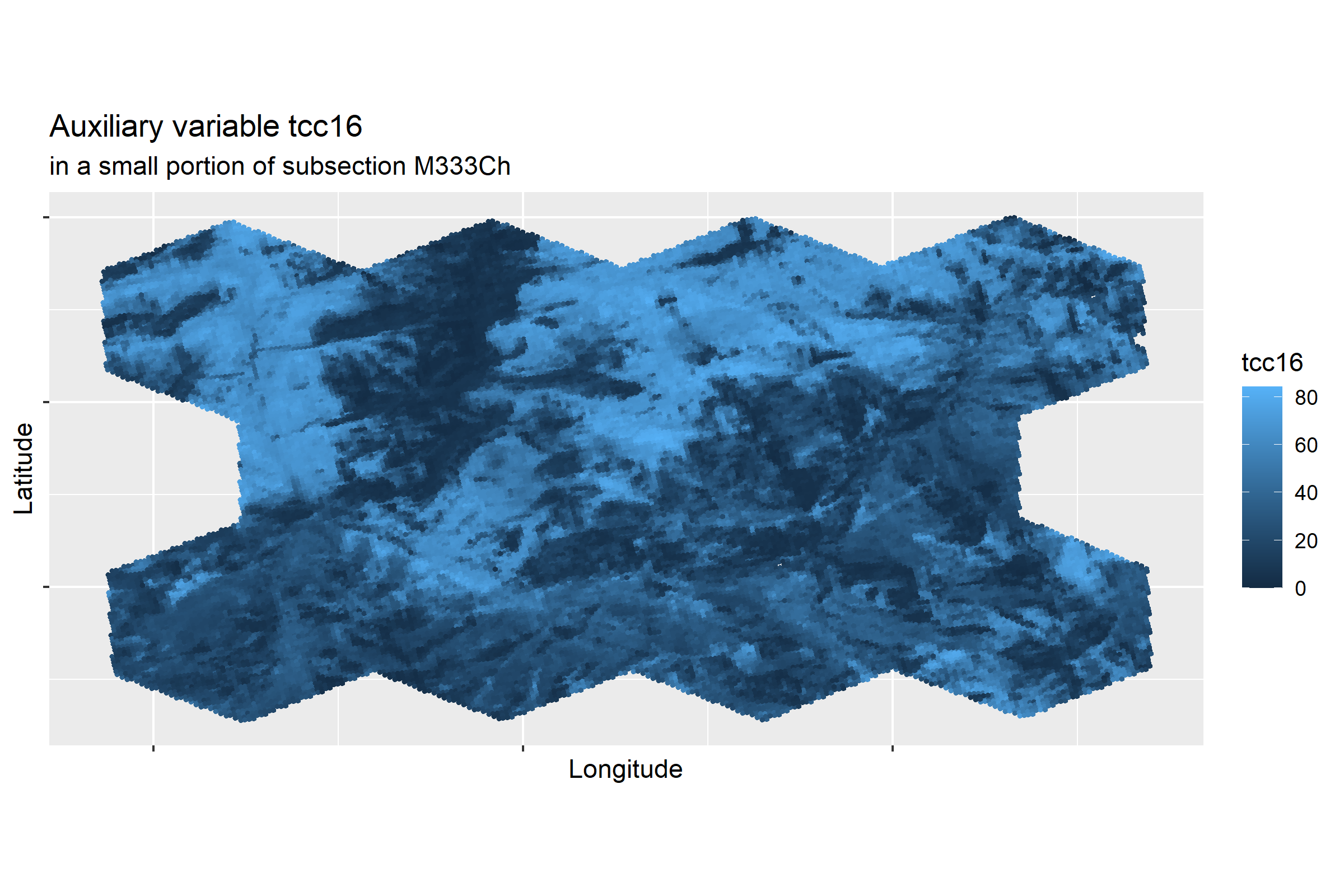}
    \caption{True values of auxiliary variable \texttt{tcc} (tree canopy cover) in a portion of subsection M333Ch.}
    \label{fig:Map_X_tcc16}
\end{figure}

Next, we checked the spatial smoothness of the imputed Y-variables, comparing kNN with various $k$ to the maps for KBAABB (Figure~\ref{fig:Map_BA_KBAABB}) and for X-variables (such as Figures~\ref{fig:Map_X_tcc16}). Spatial smoothness is not our primary goal because we are not yet attempting to fit spatial SAE models, though we may want to fit them in the future. However, we do want to have a realistic association between Ys and Xs at the unit level, and if the Xs are spatially smooth but the Ys were clearly not,
that would indicate that we probably did not have a realistic association between Ys and Xs in our artificial population.
The upper subfigure of Figure~\ref{fig:Map_BA_k1and5} shows kNN with $k$=1, which looks smoother than Figure~\ref{fig:Map_BA_KBAABB} did, in the sense that there are many sizeable spatial regions of near-constant Y-values. However, for $k$=5 (lower subfigure of Figure~\ref{fig:Map_BA_k1and5}) or greater (Figures~\ref{fig:Map_BA_k10and20} and \ref{fig:Map_BA_k50and100}), there is substantially more noise, in the sense that many pixels are quite likely to have individual spatial neighbors whose Y value is very far from their own. Although larger-scale spatial patterns can still be discerned in the plots for larger $k$, we deemed their lack of small-scale spatial smoothness to indicate unrealistic imputations.

We also checked whether the spatial smoothness of imputed Ys looks reasonable compared to the spatial smoothness of Xs, for which we have the complete population data. For example, looking at a small portion of the map (zoomed in so that we can see each individual pixel adequately), the spatial patterns in the imputed variable BA (Figure~\ref{fig:Map_BA_KBAABB}) appear to be only a little noisier than spatial patterns for related auxiliary variables such as \texttt{tcc} (Figure~\ref{fig:Map_X_tcc16}). However, note that many of the auxiliary variables have been spatially smoothed themselves, so it is not surprising that the imputed Ys do not look quite as smooth.

\newpage

\section{KBAABB implementation and sampling in our FIA setting}

\subsection{KBAABB implementation in our FIA setting}\label{sec:Implementation}

As in Algorithm~\ref*{alg:alg1}, we decided to form strata or ``donation classes'' using the categorical variable \texttt{wc2cl} or World Cover 2-level classification: a ``tree or nontree'' classification for each pixel, based on satellite measurements. By imputing separately within \texttt{wc2cl} classes, we reduce the risk of unrealistically imputing a heavily-forested pixel's Y values to an unforested pixel and vice versa.

Next, to avoid the curse of dimensionality, we relied on subject-matter expertise to choose a subset of the quantitative auxiliary variables available for matching. Eight variables were chosen so that they all measured distinct concepts and were thought to be at least somewhat predictive of the Y variables to be imputed.
As a quality check, we confirmed that the selected subset of X variables had low pairwise correlations within each \texttt{wc2cl} donation class, in the full population as well as in the survey sample. At this point, the curse of dimensionality and high variability were deemed not to be major concerns.

Next, we transformed the most heavily skewed of the selected variables to be more symmetric. Right-skewed variables were transformed as $\log(c_j + X_j)$ and left-skewed variables as $\log(c_j - X_j)$, with a separate constant $c_j$ chosen for each variable $X_j$ to ensure positive values inside the logarithm.

After this step, we also centered and scaled each of the selected variables to have mean 0 and standard deviation 1 (within each \texttt{wc2cl} class separately) on the full-population auxiliary dataset. The same centering and scaling constants (by class and variable) were then applied to the smaller survey sample dataset. This removed the effect of different variables having different units and scales, ensuring that each variable would have effectively the same weight in the kNN matching process \citep{james2021introduction}.

With these selected and transformed variables, we confirmed again that the pairwise correlations were still very small. This allowed us to use simple Euclidean distance, instead of e.g.\ Mahalanobis distance that would account for covariances between variables \citep{crookston2008yaimpute}.

The software we used to carry out the kNN matching step was the \texttt{FNN} R package \citep{li2019fnn}. For every recipient row, we match on the selected, transformed variables to find its donor pool of 10 NNs. We use the bootstrap-probability-based weights described above $(0.368^{j-1} \times 0.632)$ to sample one of each recipient's 10 possible donors. We impute all of the response Y variables from that donor to the recipient.

Finally, to define the population means for each Y variable and domain that are our targets of estimation in the simulation study, we simply take the mean of that Y variable across all of that domain's pixels in the artificial population.


\subsection{Sampling for design-based simulations in our FIA setting}\label{sec:Sampling}

After completing a run of KBAABB, our artificial population consists of around 12 million pixels,
corresponding to approximately 4000 hexagons
with nearly 3000 pixels per hexagon.

For each replication (``rep'') of a survey sample from this population, we take a sample of one pixel per hexagon, with pixels chosen uniformly at random with each hexagon. This is approximately a form of systematic sampling, since the hexagons uniformly  tile the geographic area of interest.
Then we drop any locations that happen to fall outside province M333.
This lets us carry out design-based simulations which respect the sampling design of the actual survey data. When we fit a given model form to each rep, and then summarize the model's empirical properties across reps, these properties are reasonable stand-ins for what we could expect to see under repeated sampling from the real sampling design.


\newpage

\section{Exploring KBAABB output with an R Shiny application}\label{sec:Shiny}


Alongside this paper we have created a draft R Shiny application \citep{chang2022shiny} for exploring the artificial population and a few estimators. The app can be accessed at \url{https://civilstat.shinyapps.io/fia-simpop-app/}
and the code is available at \url{https://github.com/ColbyStatSvyRsch/FIA-simpop-app}.

For each of the 100 reps currently used in the R Shiny app, on each of the 23 domains we calculate three estimators: a direct Horvitz-Thompson estimator, a post-stratified estimator,
and a small area estimator using the unit-level Battese-Harter-Fuller model.
Each model is specified on the app's ``About'' page.
Finally, for every domain and each of these three estimators, we calculate the relative bias; the $\widehat{\mbox{MSE}}/\mbox{MSE}$ ratio; and the 95\% CI coverage, as in Section~\ref{sec:CaseStudies}.

\begin{figure}
    \includegraphics[width=.9\textwidth]{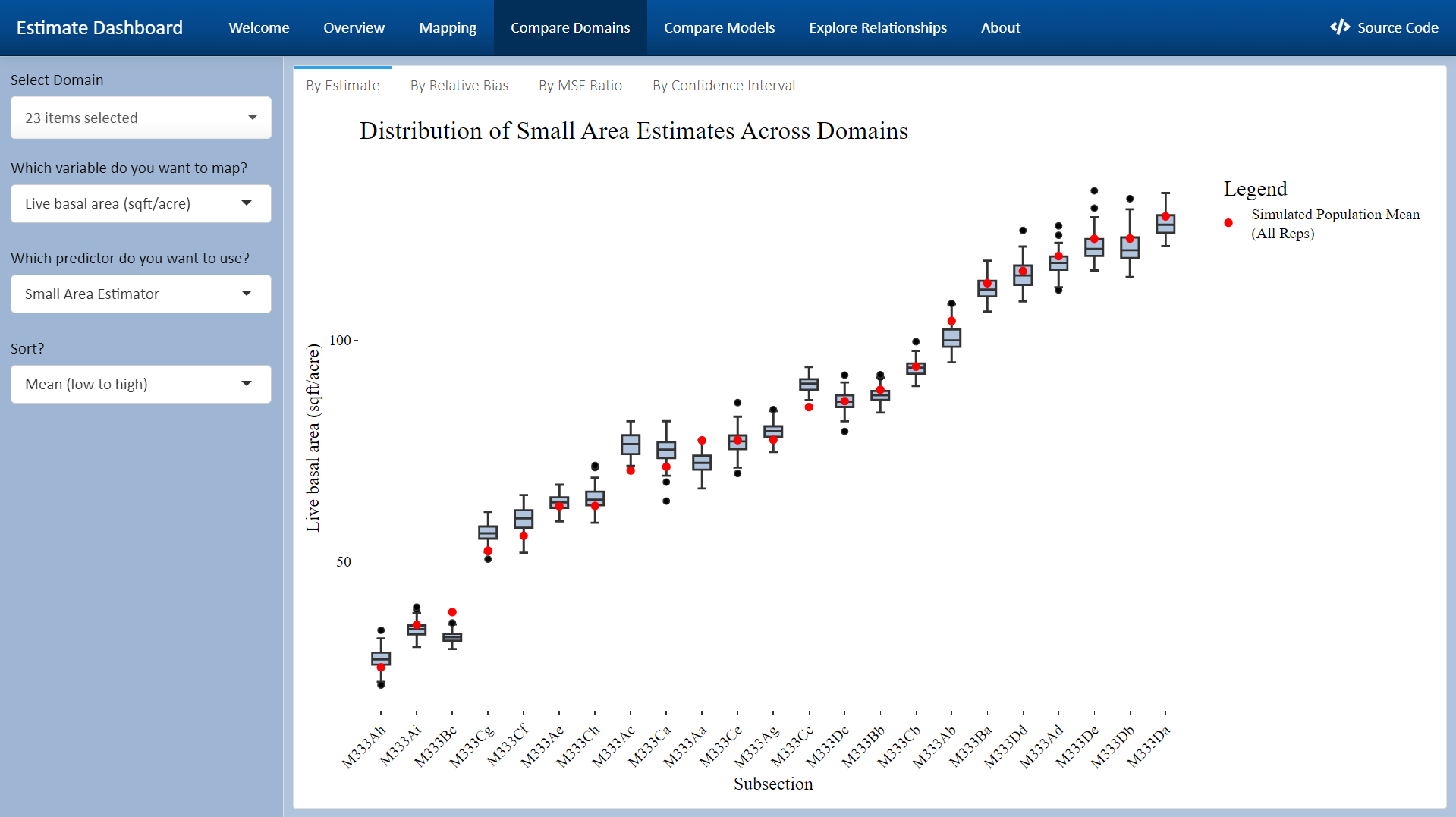}
    \caption{Screenshot of R Shiny app showing small area estimates from a model for average basal area per acre, for each of 23 domains. Each boxplot summarizes 100 reps.}
    \label{fig:BuggySAEs}
\end{figure}

\begin{figure}
    \includegraphics[width=.9\textwidth]{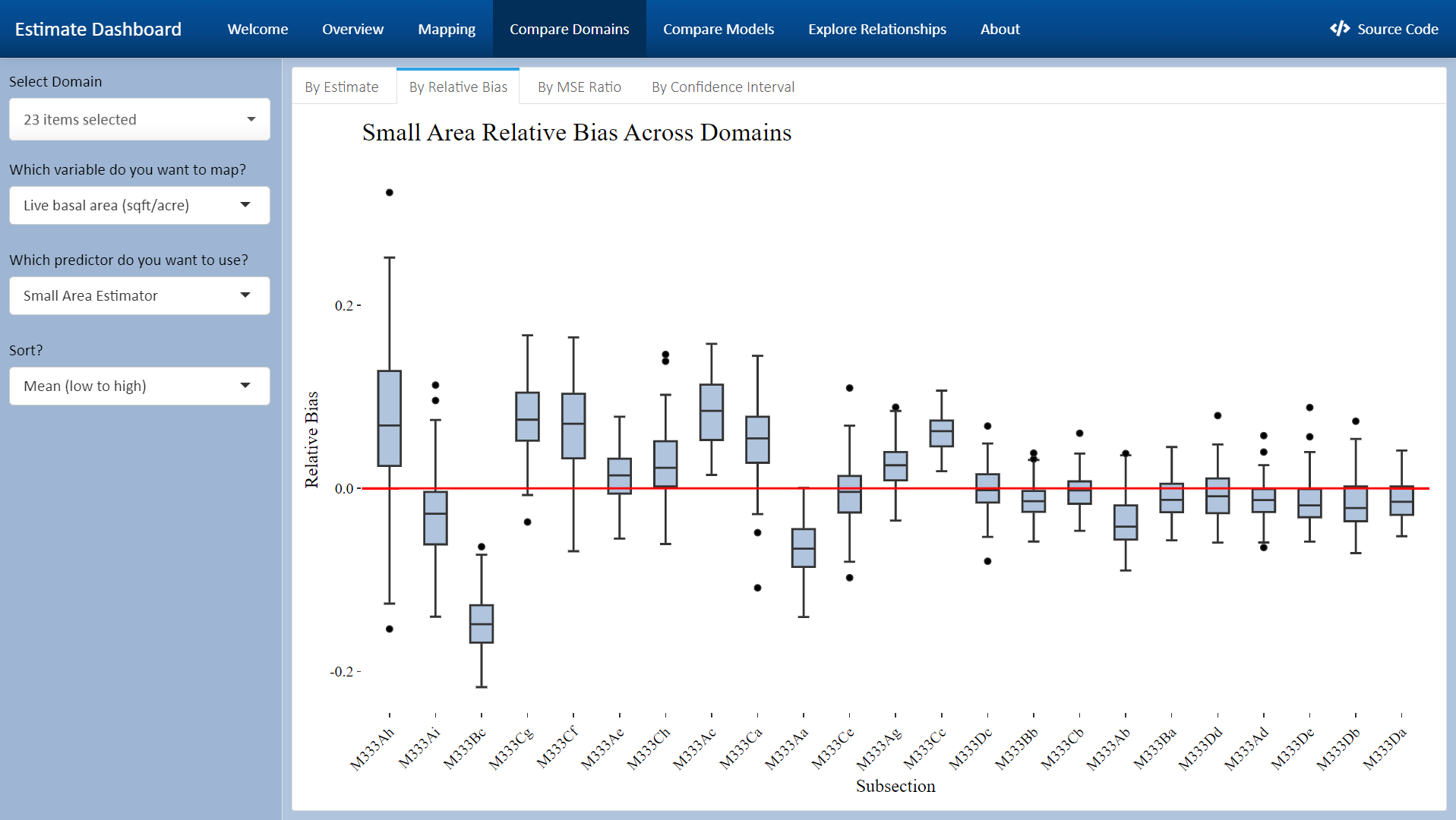}
    \caption{Screenshot of R Shiny app showing the relative biases of SAE models for average basal area per acre, for each of 23 domains. Each boxplot summarizes 100 reps.}
    \label{fig:ShinyRelBias}
\end{figure}

\begin{figure}
    \includegraphics[width=.9\textwidth]{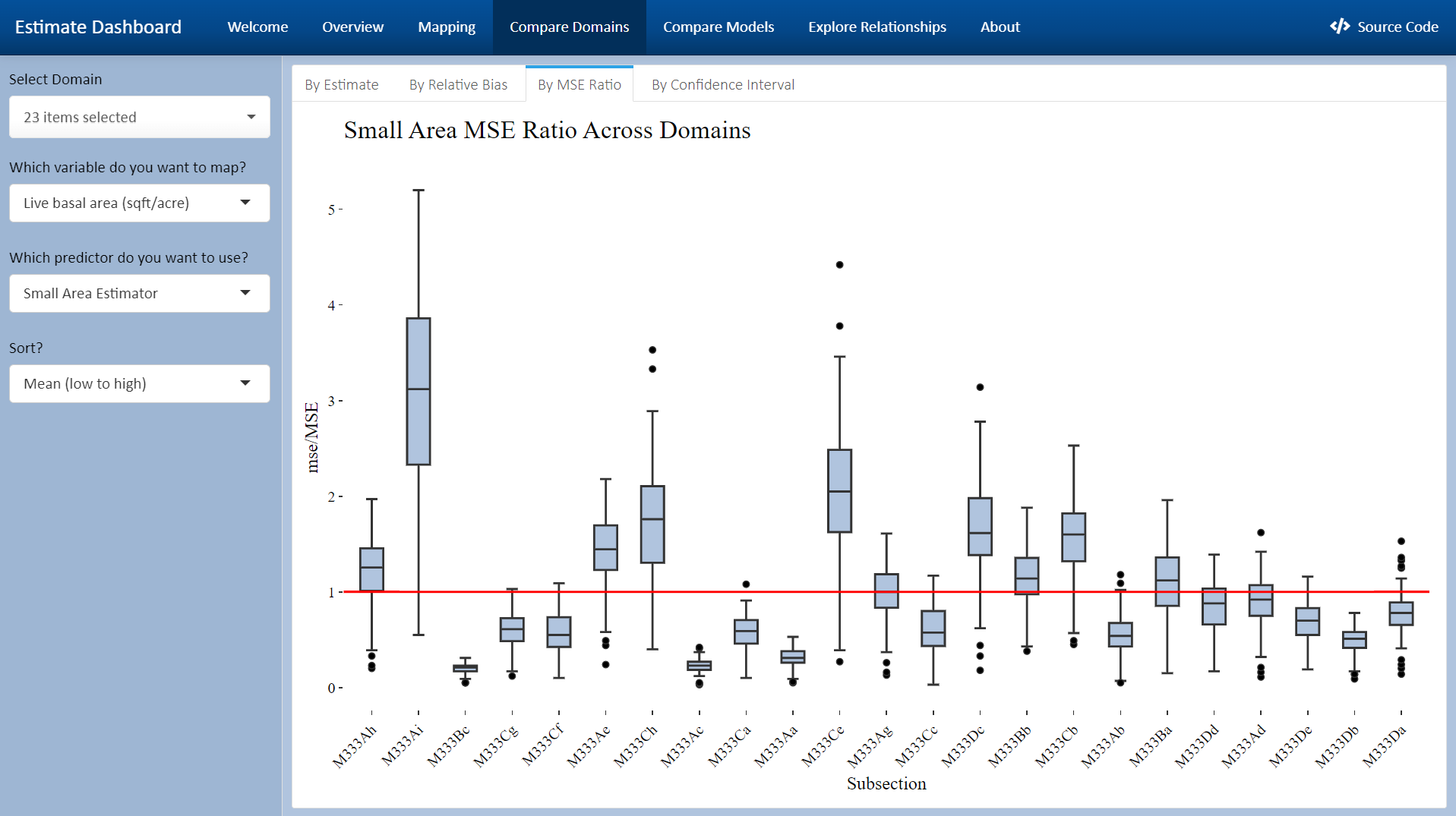}
    \caption{Screenshot of R Shiny app showing the $\widehat{\mbox{MSE}}/\mbox{MSE}$ ratios of SAE models for average basal area per acre, for each of 23 domains. Each boxplot summarizes 100 reps.}
    \label{fig:ShinyMSE}
\end{figure}

\begin{figure}
    \includegraphics[width=.9\textwidth]{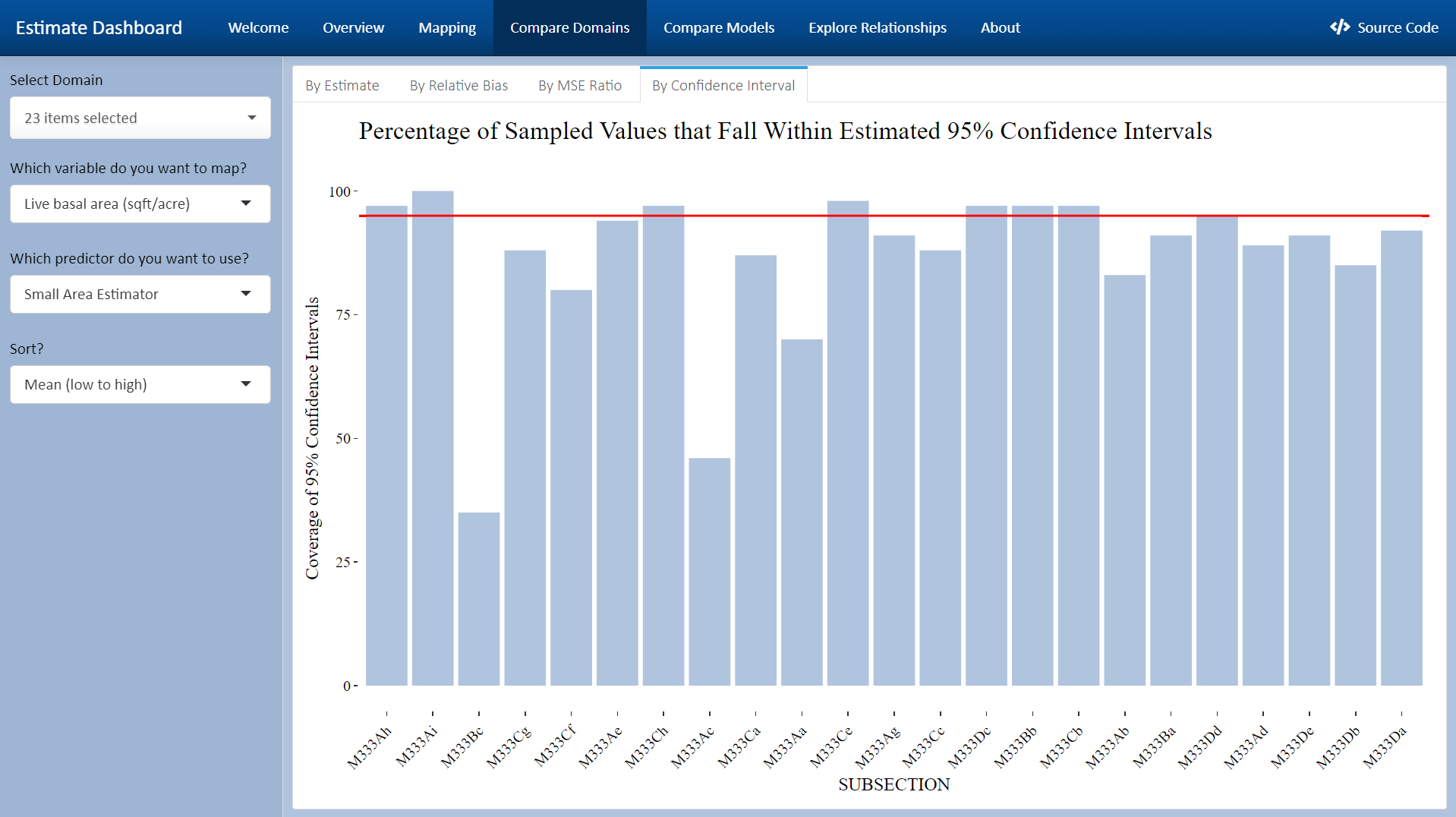}
    \caption{Screenshot of R Shiny app showing coverage of SAE models' 95\% confidence intervals for average basal area per acre, for each of 23 domains. Each bar summarizes 100 reps.}
    \label{fig:ShinyCovg}
\end{figure}

The app includes a ``Compare Domains'' page, in which the user can select one or more domains; a response variable; an estimator; and how to sort the domains. For these selections, boxplots for each of the 23 domains are used to show the distribution across 100 reps of estimates (Figure~\ref{fig:BuggySAEs}, lower subfigure), relative biases (Figure~\ref{fig:ShinyRelBias}), MSE ratios (Figure~\ref{fig:ShinyMSE}), or CI coverages (Figure~\ref{fig:ShinyCovg}) on different tabs within the page.

\begin{figure}
    \includegraphics[width=.9\textwidth]{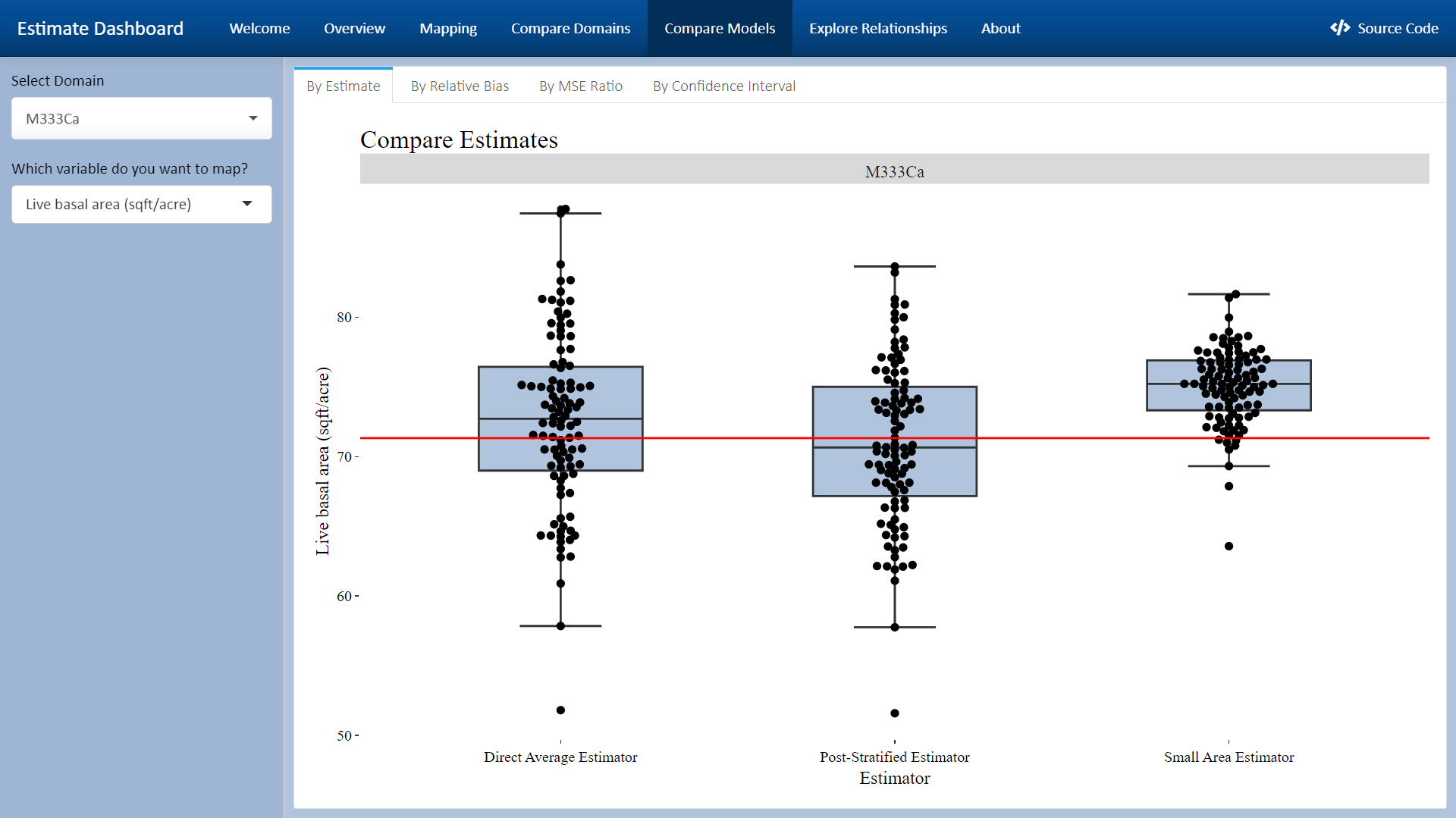}
    \caption{Screenshot of R Shiny app showing the estimates under three estimators for average basal area per acre, for one selected domain. Each boxplot summarizes 100 reps.}
    \label{fig:ShinyCompareModels}
\end{figure}

The next page is ``Compare Models'' (Figure~\ref{fig:ShinyCompareModels}), which is similar to ``Compare Domains'' except that it shows all three estimators for one domain, rather than multiple domains for one estimator.

\begin{figure}
    \includegraphics[width=.9\textwidth]{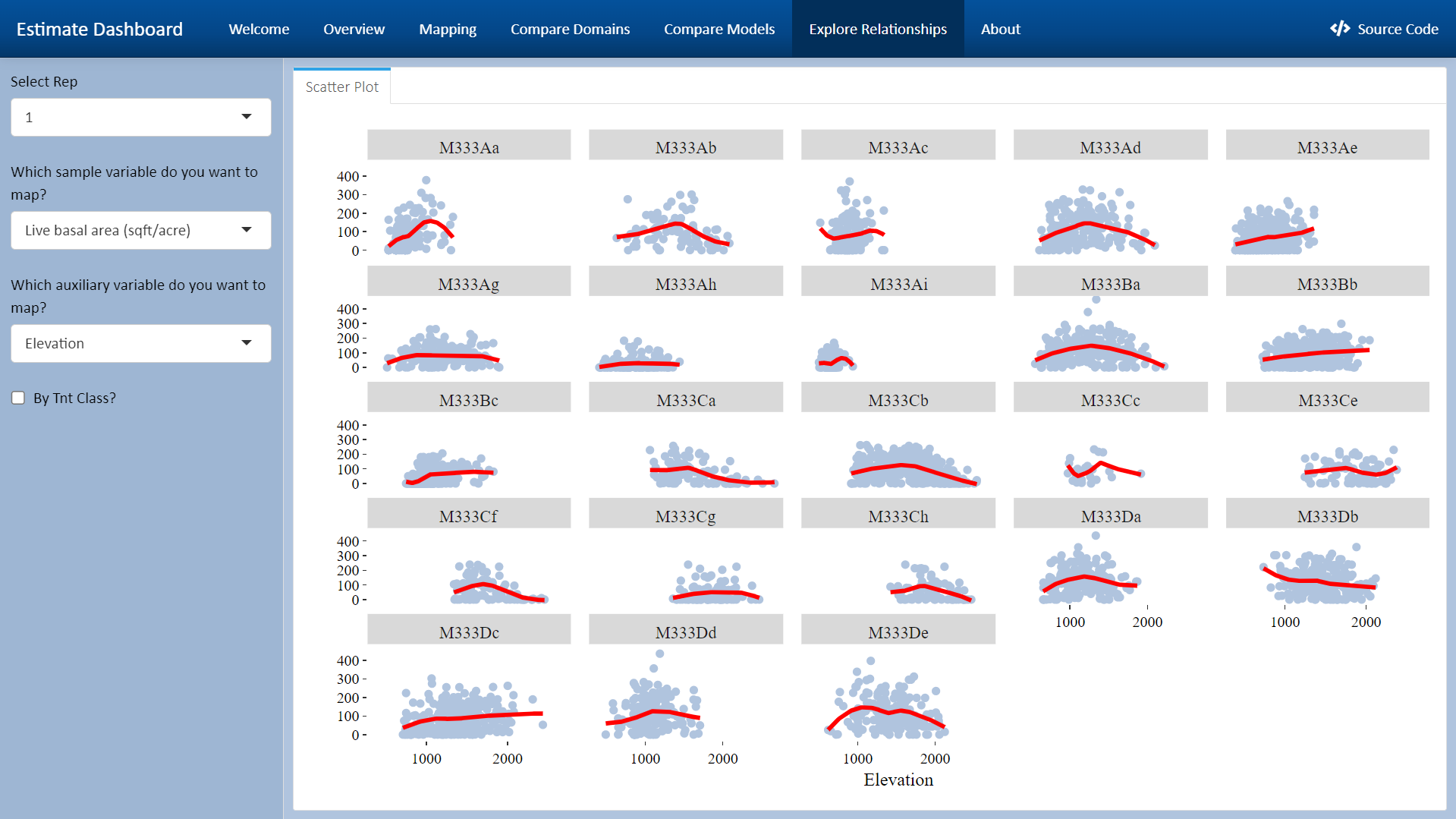}
    \caption{Screenshot of R Shiny app showing scatterplots of average basal area per acre against Elevation, for each of 23 domains, for sampled pixels in the first rep only.}
    \label{fig:ShinyExplore}
\end{figure}

Finally, the ``Explore Relationships'' page (Figure~\ref{fig:ShinyExplore}) allows the user to view scatterplots of any response variable against any auxiliary variable, faceted into small multiples---one for each domain. This can be useful for model diagnostics: are there some auxiliary variables for which the unit-level models should differ substantially across domains?

\begin{figure}
    \includegraphics[width=.9\textwidth]{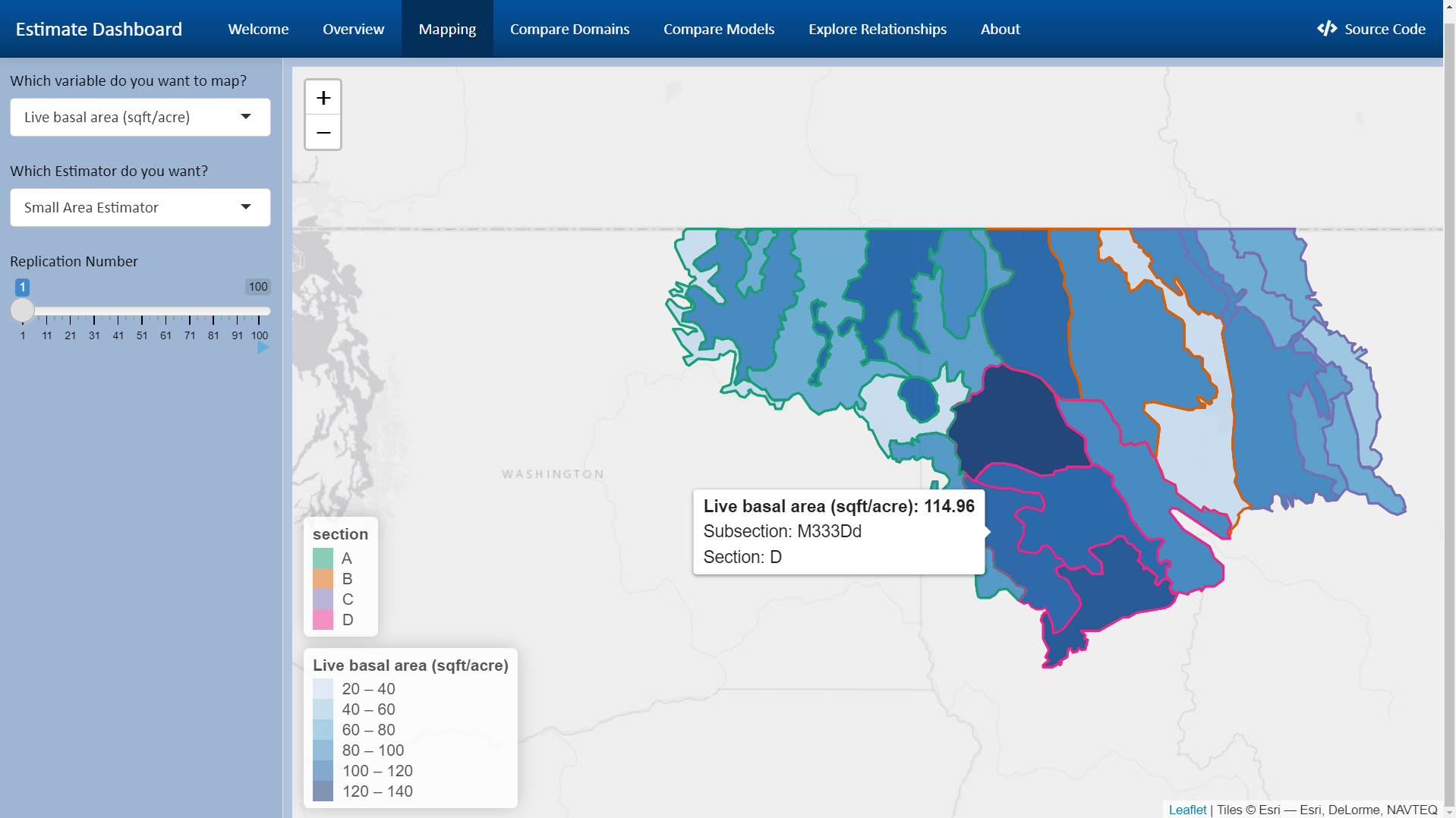}
    \caption{Screenshot of R Shiny app showing a chloropleth map of SAE estimates for average basal area per acre for all of the 23 domains. One rep is selected, but the ``Play'' button can be used to start an animation to show the map for different reps.}
    \label{fig:ShinyMapping}
\end{figure}

We have also included a Mapping page (Figure~\ref{fig:ShinyMapping}), which shows a choropleth map of each domain's estimates for a given estimator and rep. This allows users to see any spatial patterns that may be present in the estimates, as well as to explore how much these patterns vary across reps.

We are currently using this R Shiny app to help us check and compare small area models ourselves. As we develop better models and concrete advice for users of small area estimators for this population, we expect that the R Shiny app will also become useful for disseminating such advice to the diverse group of FIA users.

\printbibliography[heading=subbibliography]

@article{alfons2011simulation,
  title={Simulation of close-to-reality population data for household surveys with application to {EU-SILC}},
  author={Alfons, Andreas and Kraft, Stefan and Templ, Matthias and Filzmoser, Peter},
  journal={Statistical Methods \& Applications},
  volume={20},
  number={3},
  pages={383--407},
  year={2011},
  publisher={Springer}
}

@article{andridge2010review,
  title={A review of hot deck imputation for survey non-response},
  author={Andridge, Rebecca Roberts and Little, Roderick JA},
  journal={International Statistical Review},
  volume={78},
  number={1},
  pages={40--64},
  year={2010},
  publisher={Wiley Online Library}
}

@article{andridge2021finding,
  title={Finding a flexible hot-deck imputation method for multinomial data},
  author={Andridge, Rebecca Roberts and Bechtel, Laura and Thompson, Katherine Jenny},
  journal={Journal of Survey Statistics and Methodology},
  volume={9},
  number={4},
  pages={789--809},
  year={2021},
  publisher={Oxford University Press}
}

@article{andridge2023adapting,
  title={Adapting Nearest Neighbor for Multiple Imputation: Advantages, Challenges, and Drawbacks},
  author={Andridge, Rebecca Roberts and Thompson, Katherine Jenny},
  journal={Journal of Survey Statistics and Methodology},
  volume={11},
  issue={1},
  pages={213--233},
  year={2023}
}

@article{baffetta2009design,
  title={Design-based approach to k-nearest neighbours technique for coupling field and remotely sensed data in forest surveys},
  author={Baffetta, Federica and Fattorini, Lorenzo and Franceschi, Sara and Corona, Piermaria},
  journal={Remote Sensing of Environment},
  volume={113},
  number={3},
  pages={463--475},
  year={2009},
  publisher={Elsevier}
}

@article{battese1988error,
  title={An error-components model for prediction of county crop areas using survey and satellite data},
  author={Battese, George E and Harter, Rachel M and Fuller, Wayne A},
  journal={Journal of the American Statistical Association},
  volume={83},
  number={401},
  pages={28--36},
  year={1988},
  publisher={Taylor \& Francis}
}

@misc{bechtold2005enhanced,
  title={The {E}nhanced {F}orest {I}nventory and {A}nalysis {P}rogram---{N}ational {S}ampling {D}esign and {E}stimation {P}rocedures},
  editor={Bechtold, William A and Patterson, Paul L},
  year={2005},
  location={Gen. Tech. Report SRS-80. U.S. Department of Agriculture, Forest Service, Southern Research Station}
}

@inproceedings{brown2001evaluation,
  title={Evaluation of small area estimation methods—an application to unemployment estimates from the {UK} {LFS}},
  author={Brown, Gary and Chambers, Ray and Heady, Patrick and Heasman, Dick},
  booktitle={Proceedings of Statistics Canada Symposium},
  volume={2001},
  pages={1--10},
  year={2001},
  organization={Statistics Canada}
}

@article{burrill2021forest,
  title={The {F}orest {I}nventory and {A}nalysis {D}atabase: Database Description and User Guide version 9.0.1 for {P}hase 2. {USDA} {F}orest {S}ervice},
  author={Burrill, EA and DiTommaso, AM and Turner, JA and Pugh, SA and Christensen, G and Perry, CJ and Conkling, BL},
  url={http://www.fia.fs.fed.us/library/database-documentation/},
  year={2021}
}

@Manual{chang2022shiny,
    title = {{shiny}: Web Application Framework for {R}},
    author = {Winston Chang and Joe Cheng and JJ Allaire and Carson Sievert and Barret Schloerke and Yihui Xie and Jeff Allen and Jonathan McPherson and Alan Dipert and Barbara Borges},
    year = {2022},
    note = {R package version 1.7.3},
    url = {https://CRAN.R-project.org/package=shiny},
  }

@article{chen2000nearest,
  title={Nearest neighbor imputation for survey data},
  author={Chen, Jiahua and Shao, Jun},
  journal={Journal of Official Statistics},
  volume={16},
  number={2},
  pages={113},
  year={2000},
  publisher={Statistics Sweden (SCB)}
}

@misc{cleland2007ecological,
  title={Ecological subregions: sections and subsections of the conterminous {U}nited {S}tates, presentation scale [1:3,500,000] [{CD-ROM}]},
  author={Cleland, David T and Freeouf, Jerry A and Keys, James E and Nowacki, Greg J and Carpenter, Constance A and McNab, W Henry},
  location={A.M. Sloan, cartographer, Gen. Tech. Report WO-76. Washington, DC: U.S. Department of Agriculture, Forest Service},
  year={2007}
}

@article{crookston2008yaimpute,
  title={{yaImpute}: an {R} package for {kNN} imputation},
  author={Crookston, Nicholas L and Finley, Andrew O},
  journal={Journal of Statistical Software},
  volume={23},
  pages={1--16},
  year={2008},
  url={https://doi.org/10.18637/jss.v023.i10}
}

@article{daly2002knowledge,
  title={A knowledge-based approach to the statistical mapping of climate},
  author={Daly, Christopher and Gibson, Wayne P and Taylor, George H and Johnson, Gregory L and Pasteris, Phillip},
  journal={Climate Research},
  volume={22},
  number={2},
  pages={99--113},
  year={2002}
}

@article{dorazio2015integration,
  title={Integration and imputation of survey data in {R}: the {StatMatch} package},
  author={D’Orazio, Marcello},
  journal={Romanian Statistical Review},
  volume={63},
  number={2},
  pages={57--68},
  year={2015},
  publisher={Romanian Statistical Review}
}

@article{dorfman2018towards,
  title={Towards a routine external evaluation protocol for small area estimation},
  author={Dorfman, Alan H},
  journal={International Statistical Review},
  volume={86},
  number={2},
  pages={259--274},
  year={2018},
  publisher={Wiley Online Library}
}

@misc{ecomap1993national,
  title={National hierarchical framework of ecological units},
  author={{ECOMAP}},
  location={Unpublished administrative paper. Washington, DC: U.S. Department of Agriculture, Forest Service},
  year={1993}
}

@article{fay1979estimates,
  title={Estimates of income for small places: an application of {James-Stein} procedures to census data},
  author={Fay, Robert E and Herriot, Roger A},
  journal={Journal of the American Statistical Association},
  volume={74},
  number={366a},
  pages={269--277},
  year={1979},
  publisher={Taylor \& Francis}
}

@Manual{frescino2022fiestanalysis,
    title = {{FIESTA}nalysis: Analysis Functions for the {FIESTA} Package},
    author = {Tracey S. Frescino and Grayson W. White},
    year = {2022},
    note = {R package version 0.2.1},
  }

@Manual{frescino2023fiestautils,
title = {{FIESTAutils}: Utility Functions for Forest Inventory Estimation and Analysis},
author = {Tracey S. Frescino and Chris Toney and Grayson W. White},
year = {2023},
note = {R package version 1.1.5},
url = {https://CRAN.R-project.org/package=FIESTAutils}
}

@article{frescino2023fiesta,
title = {{FIESTA}: A Forest Inventory Estimation and Analysis {R} Package},
author = {Tracey S. Frescino and Gretchen G. Moisen and Paul L. Patterson and Chris Toney and Grayson W. White},
year = {2023},
journal = {Ecography Software Note},
doi = {10.1111/ecog.06428}
}

@inproceedings{heath2009investigation,
  title={Investigation into calculating tree biomass and carbon in the {FIADB} using a biomass expansion factor approach},
  author={Heath, Linda S and Hansen, Mark and Smith, James E and Miles, Patrick D and Smith, Brad W},
  booktitle={Proceedings of the Forest Inventory and Analysis (FIA) Symposium 2008},
  pages={21--23},
  year={2009}
}

@article{horvitz52HT,
         author = {Horvitz, D. G. and Thompson, D. J.},
         title = {A generalization of sampling without replacement
		  from a finite universe},
         journal = {Journal of the American Statistical
		  Association},
         volume=47,
         year = 1952,
         pages = {663-685}
     }

@article{isaki1982survey,
  title={Survey design under the regression superpopulation model},
  author={Isaki, Cary T and Fuller, Wayne A},
  journal={Journal of the American Statistical Association},
  volume={77},
  number={377},
  pages={89--96},
  year={1982},
  publisher={Taylor \& Francis}
}

@book{james2021introduction,
  title={An Introduction to Statistical Learning with Applications in {R}},
  author={James, Gareth and Witten, Daniela and Hastie, Trevor and Tibshirani, Robert},
  year={2021},
  edition={2},
  publisher={Springer}
}

@incollection{lehtonen2009design,
  title={Design-based methods of estimation for domains and small areas},
  author={Lehtonen, Risto and Veijanen, Ari},
  booktitle={Handbook of Statistics},
  volume={29B},
  chapter={31},
  pages={219--249},
  year={2009},
  publisher={Elsevier}
}

@Manual{li2019fnn,
title = {{FNN}: Fast Nearest Neighbor Search Algorithms and Applications},
author = {Li, Shengqiao},
year = {2019},
note = {R package version 1.1.3.1},
url = {https://CRAN.R-project.org/package=FNN},
}

@article{mcroberts2005enhanced,
  title={The enhanced {F}orest {I}nventory and {A}nalysis program of the {USDA} {F}orest {S}ervice: Historical perspective and announcement of statistical documentation},
  author={McRoberts, Ronald E and Bechtold, William A and Patterson, Paul L and Scott, Charles T and Reams, Gregory A},
  journal={Journal of Forestry},
  volume={103},
  number={6},
  pages={304--308},
  year={2005},
  publisher={Oxford University Press}
}

@article{morris2014tuning,
  title={Tuning multiple imputation by predictive mean matching and local residual draws},
  author={Morris, Tim P and White, Ian R and Royston, Patrick},
  journal={BMC Medical Research Methodology},
  volume={14},
  number={75},
  pages={1--13},
  year={2014},
  publisher={Springer}
}

@article{nur2005dealing,
  title={Dealing with incomplete data in questionnaires of food and alcohol consumption},
  author={Nur, U A M and Longford, N T and Cade, J E and Greenwood, D C},
  journal={Statistics in Transition},
  volume={7},
  number={1},
  pages={111--134},
  year={2005},
  publisher={Polish Statistical Association}
}

@article{picotte2019landfire,
  title={{LANDFIRE} remap prototype mapping effort: Developing a new framework for mapping vegetation classification, change, and structure},
  author={Picotte, Joshua J and Dockter, Daryn and Long, Jordan and Tolk, Brian and Davidson, Anne and Peterson, Birgit},
  journal={Fire},
  volume={2},
  number={2},
  pages={35},
  year={2019},
  publisher={MDPI}
}

@book{rao15small,
  title={Small Area Estimation},
  author={Rao, John N K and Molina, Isabel},
  edition={2},
  year={2015},
  publisher={John Wiley \& Sons}
}

@article{rollins2009landfire,
  title={{LANDFIRE}: a nationally consistent vegetation, wildland fire, and fuel assessment},
  author={Rollins, Matthew G},
  journal={International Journal of Wildland Fire},
  volume={18},
  number={3},
  pages={235--249},
  year={2009},
  publisher={CSIRO Publishing}
}

@article{rubin1986multiple,
  title={Multiple imputation for interval estimation from simple random samples with ignorable nonresponse},
  author={Rubin, Donald B and Schenker, Nathaniel},
  journal={Journal of the American Statistical Association},
  volume={81},
  number={394},
  pages={366--374},
  year={1986},
  publisher={Taylor \& Francis}
}

@book{rubin1987multiple,
  title={Multiple imputation for nonresponse in surveys},
  author={Rubin, Donald B},
  year={2004},
  publisher={John Wiley \& Sons}
}

@article{schwager2021remote,
  title={Remote sensing variables improve species distribution models for alpine plant species},
  author={Schwager, Patrick and Berg, Christian},
  journal={Basic and Applied Ecology},
  volume={54},
  pages={1--13},
  year={2021},
  publisher={Elsevier}
}

@article{templ2017simulation,
  title={Simulation of synthetic complex data: The {R} package {simPop}},
  author={Templ, Matthias and Meindl, Bernhard and Kowarik, Alexander and Dupriez, Olivier},
  journal={Journal of Statistical Software},
  volume={79},
  number={10},
  pages={1--38},
  year={2017},
  publisher={UCLA, Dept. of Statistics}
}

@article{tzavidis2018start,
  title={From start to finish: a framework for the production of small area official statistics},
  author={Tzavidis, Nikos and Zhang, Li-Chun and Luna, Angela and Schmid, Timo and Rojas-Perilla, Natalia},
  journal={Journal of the Royal Statistical Society, Series A (Statistics in Society)},
  volume={181},
  number={4},
  pages={927--979},
  year={2018}
}

@misc{usda2014farm,
    title={Agricultural {A}ct of 2014}, 
    url={https://www.congress.gov/113/plaws/publ79/PLAW-113publ79.pdf},
    location={H.R.2642, 113th Congress (2013-2014)}, 
    year={2014}
}

@misc{usda2018farm,
    title={Agriculture {I}mprovement {A}ct of 2018}, 
    url={https://www.agriculture.senate.gov/imo/media/doc/Agriculture%20Improvement%20Act%20of%202018.pdf},
    location={H.R.2, 115th Congress (2017-2018)}, 
    year={2018}
}

@misc{usgs2019ned,
    author={{U.S. Geological Survey}}, 
    title={{LANDFIRE} {E}levation}, 
    location={USGS EROS, Sioux Falls, South Dakota}, 
    year={2019}
}

@inproceedings{wieczorek2012bayesian,
  title={A {B}ayesian zero-one inflated beta model for small area shrinkage estimation},
  author={Wieczorek, Jerzy and Nugent, Ciara and Hawala, Sam},
  booktitle={Proceedings of the 2012 Joint Statistical Meetings, American Statistical Association, Alexandria, VA},
  year={2012}
}

@inproceedings{wieczorek2013empirical,
  title={An Empirical Artificial Population and Sampling Design for Small-Area Model Evaluation},
  author={Wieczorek, Jerzy and Franco, Carolina},
  booktitle={Proceedings of the 2013 Joint Statistical Meetings, American Statistical Association, Alexandria, VA},
  year={2013}
}

@article{wojcik2022gregory,
  title={{GREGORY}: A Modified Generalized Regression Estimator Approach to Estimating Forest Attributes in the Interior Western {US}},
  author={Wojcik, Olek C and Olson, Samuel D and Nguyen, Paul-Hieu V and McConville, Kelly S and Moisen, Gretchen G and Frescino, Tracey S},
  journal={Frontiers in Forests and Global Change},
  volume={4},
  year={2022},
  publisher={Frontiers}
}

@article{woodruff1966use,
  title={Use of a regression technique to produce area breakdowns of the monthly national estimates of retail trade},
  author={Woodruff, Ralph S},
  journal={Journal of the American Statistical Association},
  volume={61},
  number={314},
  pages={496--504},
  year={1966},
  publisher={Taylor \& Francis}
}

@article{yang2018new,
  title={A new generation of the {U}nited {S}tates {N}ational {L}and {C}over {D}atabase: Requirements, research priorities, design, and implementation strategies},
  author={Yang, Limin and Jin, Suming and Danielson, Patrick and Homer, Collin and Gass, Leila and Bender, Stacie M and Case, Adam and Costello, Catherine and Dewitz, Jon and Fry, Joyce and Funk, Michelle and Granneman, Brian and Liknes, Greg C and Rigge, Matthew and Xian, George},
  journal={ISPRS Journal of Photogrammetry and Remote Sensing},
  volume={146},
  pages={108--123},
  year={2018},
  publisher={Elsevier}
}

@incollection{yang2019nearest,
  title={Nearest neighbor imputation for general parameter estimation in survey sampling},
  author={Yang, Shu and Kim, Jae Kwang},
  booktitle={The Econometrics of Complex Survey Data},
  volume={39},
  pages={209--234},
  year={2019},
  publisher={Emerald Publishing Limited}
}

@article{zanaga2021esa,
  title={{ESA} {WorldCover} 10 m 2020 v100},
  author={Zanaga, Daniele and Van De Kerchove, Ruben and De Keersmaecker, W and Souverijns, N and Brockmann, Carsten and Quast, R and Wevers, Jan and Grosu, A and Paccini, A and Vergnaud, S and Cartus, Oliver and Santoro, Maurizio and Fritz, Steffen and Georgieva, I and Lesiv, M and Carter, S and Herold, M and Li, L and Tsendbazar, NE and Ramoino, F and Arino, O},
  year={2021},
  publisher={Zenodo}
}

@article{finley2011hierarchical,
  title={A hierarchical model for quantifying forest variables over large heterogeneous landscapes with uncertain forest areas},
  author={Finley, Andrew O and Banerjee, Sudipto and MacFarlane, David W},
  journal={Journal of the American Statistical Association},
  volume={106},
  number={493},
  pages={31--48},
  year={2011},
  publisher={Taylor \& Francis}
}

@article{white2024small,
author = {White, Grayson W. and Yamamoto, Josh K. and Elsyad, Dinan H. and Schmitt, Julian F. and Korsgaard, Niels H. and Hu, Jie Kate and Gaines, George C. and Frescino, Tracey S. and McConville, Kelly S.},
title = {Small area estimation of forest biomass via a two-stage model for continuous zero-inflated data},
journal = {Canadian Journal of Forest Research},
volume = {55},
number = {},
pages = {1-19},
year = {2025},
doi = {10.1139/cjfr-2024-0149},
URL = {https://doi.org/10.1139/cjfr-2024-0149},
eprint = {https://doi.org/10.1139/cjfr-2024-0149}
}
\end{refsection}

\end{document}